\documentclass[fleqn,usenatbib]{mnras}
\usepackage{amsmath}    
\usepackage{txfonts}
\usepackage[T1]{fontenc}
\usepackage{ae,aecompl}
\usepackage{graphicx}   
\usepackage{amssymb}    
\usepackage{caption}
\usepackage{subcaption}
\urldef{\footurl}\url{http://www.this-is-a-very-very-very-very-very-very-very-very-very-very-very-very-very-very-very-very-very-very-very-very-very-very-very-very-long.url/}
\usepackage{float}

\title[Multi-epoch \textit{Chandra} HETG study of outflow in NGC 4051]{A deep, multi-epoch \textit{Chandra} HETG study of the ionized outflow from NGC 4051}
\author[A. Ogorzalek]{
A. Ogorzalek,$^{1,2,3,4}$\thanks{E-mail: anna.ogorzalek@nasa.gov}
A. L. King,$^{5}$
S. W. Allen,$^{3,4,6}$
J. C. Raymond,$^{7}$
\& D. R. Wilkins$^{3,4}$
\\
$^{1}$NASA Goddard Space Flight Center, Code 662, Greenbelt, MD 20771, USA\\
$^{2}$1 Department of Astronomy, University of Maryland, College Park, MD 20742\\
$^{3}$Department of Physics, Stanford University, 382 Via Pueblo Mall, Stanford, CA 94305, USA\\
$^{4}$Kavli Institute for Particle Astrophysics \& Cosmology, P. O. Box 2450, Stanford University, Stanford, CA 94305, USA\\
$^{5}$Swell Energy, 1515 7th St. \#049, Santa Monica, CA 90401, USA\\
$^{6}$SLAC National Accelerator Laboratory, Menlo Park, CA 94025, USA\\
$^{7}$Center for Astrophysics, Harvard \& Smithsonian, 60 Garden St., Cambridge, MA 02138, USA
}
\date{Accepted August 15th 2022. Received YYY; in original form ZZZ}

\pubyear{2022}

\begin{document}
\label{firstpage}
\pagerange{\pageref{firstpage}--\pageref{lastpage}}

\maketitle
\begin{abstract}
Actively accreting supermassive black holes significantly impact the evolution of their host galaxies, truncating further star formation by expelling large fractions of gas with wide-angle outflows. The X-ray band is key to understanding how these black hole winds affect their environment, as the outflows have high temperatures ($\sim$10$^{5-8}$K). We have developed a Bayesian framework for characterizing Active Galactic Nuclei (AGN) outflows with an improved ability to explore parameter space and perform robust model selection.  We  applied this framework to a new 700~ks and an archival 315~ks  \textit{Chandra} High Energy Transmission Gratings observation of the Seyfert galaxy NGC~4051. We have  detected six absorbers intrinsic to NGC~4051. These wind components span velocities from 400~km~s$^{-1}$ to 30,000~km~s$^{-1}$.  We have determined that the most statistically signiﬁcant wind component is purely collisionally ionized, which is the first detection of such an absorber. This wind has $T\approx10^7$~K and $v\approx880$~km~s$^{-1}$ and remains remarkably stable between the two epochs. Other slow components also remain stable across time. Fast outflow components change their properties between 2008 and 2016, suggesting either physical changes or clouds moving in and out of the line of sight.  For one of the fast components we obtain one of the tightest wind density measurements to date, log~$n/$[cm$^{-3}$]=13.0$^{+0.01}_{-0.02}$,  and determine that it is located at  $\sim$240 gravitational radii. The estimated total outflow power surpasses 5\% of the bolometric luminosity (albeit with large uncertainties) making it important in the context of galaxy--black hole interactions. 
\end{abstract}
\begin{keywords}
galaxies: active -- galaxies: Seyfert -- quasars: absorption lines -- techniques: spectroscopic -- methods: statistical  -- X-rays: general
\end{keywords}

\section{Introduction}\label{sec:intro}

Ionized outflows from Active Galactic Nuclei (AGN) are thought to play an important role in the evolution supermassive black holes (SMBH) and their host galaxies \citep{Veilleux2005, Fabian2012, King2015}. They can both encourage black hole growth, by removing angular momentum from the gas in the accretion disk, and impede it, carrying away up to 1000 times the amount of accreting mass \citep[e.g.][]{Crenshaw2012}. During the quasar phase in the early Universe, at the peak of cosmic AGN activity,  wide-angle winds are thought to have helped shape large scale structure formation \citep{Scannapieco2004, DiMatteo2005}, and drive  the metal enrichment of the Intragalactic Medium \citep{Hamann1999, Khalatyan2008, 2013Natur.502..656W, 2017MNRAS.470.4583U, 2017MNRAS.468..531B}. These outflows may also be responsible for establishing the $M-\sigma$ relation \citep[e.g.][]{Silk1998,  King2015, Zubovas2019}, the empirical correlation between the stellar bulge velocity dispersion and SMBH mass, \citep{Ferrarese2000, Gebhardt2000, Kormendy2013}, by removing the gas from the bulge and/or halo, and quenching star formation \citep{Hopkins2005, Heckman2014, Fiore2017}.

However, how exactly these winds are launched and what energetic impact they have on the host galaxy is still unclear. Understanding this requires precise knowledge of the outflow locations and kinetic powers, which can be derived from directly observable wind properties: ionization, density, and velocity. These key physical characteristics of the outflows can be probed using absorption features the winds imprint onto the continuum radiation from the black hole accretion disk, with  higher photoionization states originating closer to the black holes (given similar densities). Studies in the X-ray band are therefore crucial for characterizing AGN wind properties by probing the gas in the immediate environment of the central engine. 

Since current X-ray spectrographs have limited spectral resolution and low effective area, accurate measurements of wind physical properties are possible only for local AGN. These are  Seyfert galaxies, which are generally low mass analogs of quasars in the local Universe. Due to their smaller size, their accretion disk emission peaks at shorter wavelengths compared to quasars, resulting in higher X-ray fluxes for comparable Eddington fractions.   Approximately  $\sim$50\%--70\% of Seyfert galaxies show evidence for outflows via detected absorption features in the UV and X-ray bands \citep{Crenshaw2003}. Their X-ray spectra are particularly rich in information, nominally providing access to many ionization states.

However, even for objects with deep X-ray spectra, physical modelling poses challenges. Typically, a grid of plasma absorption models is built using spectral synthesis codes such as XSTAR  \citep{Kallman2001} or Cloudy \citep{Ferland2017}. At this step, a photoionizing continuum is assumed and fixed (though some authors will iterate the continuum a few times, e.g. \citealt{McKernan2007}). Then, a `best fit' is found with optimization algorithms implemented in spectral analysis packages like ISIS or XSPEC. Due to computing limitations, this grid approach usually narrows down the explored parameter space by fixing or reducing the parameter range searched (e.g. the intrinsic absorption line width or outflow density). Additionally, the grids are typically sparsely populated, requiring substantial interpolation that may further bias the results.

Recently, a number of improvements have been introduced. Larger grids can be calculated with the use of high performance computer clusters \citep{Danehkar2018mpi}, and Monte Carlo Markov Chain (MCMC) methods have been explored for grid fitting \citep[e.g.][]{Danehkar2018}. Further, some models, e.g. \texttt{pion} from the spectral package SPEX \citep{SPEX}, can vary the ionizing continuum and calculate the ionization balance at each step of the fitting process \citep{Mehdipour2016}. 

In this paper, we build upon these techniques and  introduce a Bayesian framework for fitting X-ray spectra to better study AGN winds. We build our model of the AGN continuum and absorption by utilizing the hypothesis testing Divergence Information Criterion at each step \citep[DIC;][]{Spiegelhalter2002}. Our spectral fits are performed in two stages. We first  pre-search the parameter space using one of the largest-to-date grids of photoionized models. Having identified potential maxima in the likelihood space, we explore them in a self-consistent fashion, varying the ionizing continuum and calculating the wind absorption at each step of the fitting procedure. 

We apply our approach to a new 700~ks \emph{Chandra} High Energy Transmission Grating \citep[HETG;][]{Canizares2005} observation of NLSy1 NGC~4051, and an archival 315~ks observation from 2008. At a distance of 17.6~Mpc \citep{2014ApJ...784L..11Y} and a redshift of 0.002336 \citep{Verheijen2001}, NGC~4051 hosts a SMBH of mass $\sim1.3\cdot10^{6}$~M$_{\sun}$  \citep{Bentz2015} and its bolometric luminosity is 10$^{43.4}$~erg~s$^{-1}$  \citep[][]{2005A&A...431..111B}. X-ray spectra of  this source are rich in absorption from many ionic species, and the source has previously been studied with \textit{Chandra}'s Low Energy Transmission Grating \citep[LETG; ][]{Steenbrugge2009}, HETG \citep{Collinge2001,McKernan2007, Lobban2011, King2012}, and XMM-\textit{Newton}'s Reflection Grating Spectrometer  \citep[RGS; ][]{Krongold2007, Nucita2010, Pounds2011, Pounds2013, Silva2016, Mizumoto2017}.  

The structure of this paper is as follows. We describe the observations used and the data reduction in Section~\ref{sec:data}. In Section~\ref{sec:methods} we give an overview of the models used and our fitting approach. We show results of the spectral fits in Section~\ref{sec:results},  discuss systematic uncertainties in Section~\ref{sec:uncertainties} and the implications of our results in Section~\ref{sec:discussion}. A summary of the paper and conclusions are given in Section~\ref{sec:conclusions}. 

Unless otherwise stated, all error bars are 68\% credible intervals. We adopt a velocity sign convention such that negative redshifts translate to positive velocities (i.e. material moving towards the observer). 

\section{Data}\label{sec:data}

\subsection{Observations}

\begin{table}
\centering
\caption{\emph{Chandra} observations used in this work. \label{tab:obsid}}

\begin{tabular}{@{ }c@{  \: }c@{ \: }c@{ \: }c@{ \: }} 
\hline
\hline
obsID & exposure  &date  & \emph{Chandra} \\
 & [ks] & [year] & detector \\ 
\hline
\hline
\multicolumn{4}{c}{2008: 314 ks total}\\
\hline

10777   &   27.84   &   2008.850553 &   ACIS-S/HETG \\
10775   &   30.88   &   2008.855931 &   ACIS-S/HETG \\
10403   &   38.15   &   2008.859358 &   ACIS-S/HETG \\
10778   &   34.13   &   2008.862556 &   ACIS-S/HETG \\
10776   &   25.15   &   2008.865125 &   ACIS-S/HETG \\
10404   &   20.09   &   2008.867169 &   ACIS-S/HETG \\
10801   &   26.14   &   2008.869638 &   ACIS-S/HETG \\
10779   &   27.75   &   2008.888002 &   ACIS-S/HETG \\
10780   &   26.43   &   2008.901975 &   ACIS-S/HETG \\
10781   &   24.13   &   2008.905991 &   ACIS-S/HETG \\
10782   &   23.65   &   2008.913118 &   ACIS-S/HETG \\
10824   &   9.15    &   2008.915035 &   ACIS-S/HETG \\

\hline
\multicolumn{4}{c}{2016: 701 ks total}\\
\hline
18768   &   93.46   &   2016.112952 &   ACIS-S/HETG \\
17104   &   60.08   &   2016.127234 &   ACIS-S/HETG \\
17105   &   116.09  &   2016.133229 &   ACIS-S/HETG \\
18769   &   71.89   &   2016.141393 &   ACIS-S/HETG \\
17102   &   114.8   &   2016.155752 &   ACIS-S/HETG \\
18785   &   66.29   &   2016.160842 &   ACIS-S/HETG \\
17103   &   63.48   &   2016.164552 &   ACIS-S/HETG \\
18786   &   66.04   &   2016.168153 &   ACIS-S/HETG \\
18787   &   25.79   &   2016.285771 &   ACIS-S/HETG \\
18823   &   23.07   &   2016.288794 &   ACIS-S/HETG \\
\hline
\hline
\end{tabular}
\end{table}

NGC~4051 was observed by \textit{Chandra} HETG for 314~ks in 2008 and 701~ks in 2016. Observations used in this work are listed in Table~\ref{tab:obsid}. Each data set consists of the Medium Energy Grating (MEG; nominal energy range 0.4--5.0~keV) and the High Energy Grating (HEG; 0.8-10.0~keV) spectra. The 2008 data set has 77~k and 186~k counts in MEG and HEG respectively, while the 2016 data set has 120~k and 256~k. Despite a factor of two in exposure time difference, both observations have comparable number of soft photons ($<1$~keV). This is due to the  progressing soft contamination of the \textit{Chandra} ACIS chips \citep{Plucinsky2018}.

\subsection{Data reduction}

We performed standard reprocessing of the  2016 and archival 2008 observations using the CIAO (v. 4.10)  tool \textit{chandra\_repro}. The default spectral extraction regions were decreased by fifty percent to avoid overlap of the HEG by the MEG extraction above 8~keV.

The flux remained fairly constant throughout the exposures, with the average count rate varying by less than 10\% within the 2016 and 2008 data sets respectively. Therefore, we co-added all epochs within each year using the CIAO tool {\it combine\_grating\_spectra}, which also combined the positive and negative first spectral orders. In this way, we obtained HEG and MEG first order spectra. We used the HEASoft (v. 6.23) ftool {\it grppha} to ensure that the final spectra had at least one count per bin.

\section{Methods}\label{sec:methods}

In order to accurately extract information from the high resolution high S/N HETG observations of NGC~4051, we developed a Bayesian framework, wherein we construct our model by rigorously testing for the presence of physically motivated spectral components. Details of the model components considered and the assumptions that accompany them are described in Section~\ref{sec:spmodels}. We summarize how the best fit for each model considered is found in Section.~\ref{sec:mcmc}, and how we discriminate between models in Section~\ref{subsec:dic}. 
\begin{figure}
\centering
\includegraphics[width=\columnwidth,trim={0mm 0mm 0mm 0mm},clip]{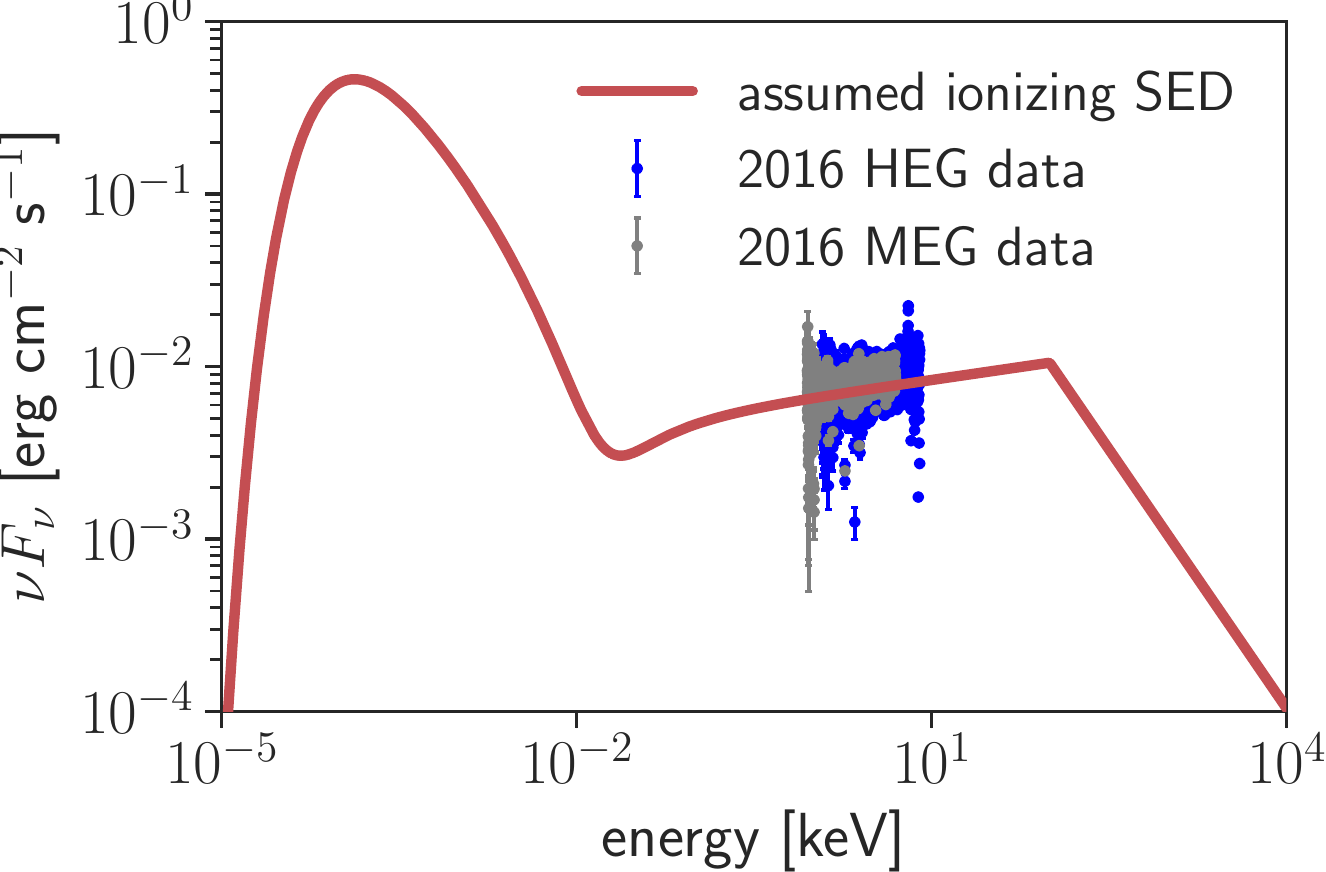} 
\caption{The SED assumed in the Cloudy photoionization calculations is shown in red, with the  2016 HEG data shown as blue data points and 2016 MEG data shown in grey. Note that the slope of the X-ray power-law is varied along with the continuum components in our fits. }
\label{fig:SED}
\end{figure}

\subsection{Spectral models} \label{sec:spmodels}

Unless otherwise noted, all models are part of the XSPEC spectral fitting package, ver.~12.10.1f \citep{Arnaud1996}. For these models, we use the solar abundances of \cite{Lodders2003}. Absorption  tables are calculated with the Cloudy spectral synthesis code ver.~17.00 \citep{Ferland2017}, for which the default Cloudy solar abundances are used. 

\subsubsection{Continuum} \label{subsec:cont}

The minimal physically motivated components of an AGN continuum  are a coronal power-law, soft excess emission, and disk reflection \citep[see e.g.][]{2013MNRAS.435.1511A}. The latter can originate from the neutral matter in the outer regions of the central engine, or from the highly ionized gas close to the black hole, where relativistic effects are important. We model the coronal emission with a \textit{powerlaw} model, and the neutral reflection with \emph{xillver} \citep{xillver1,xillver2}. 

Due to the lack of consensus on the origin of the soft excess,  we consider both the commonly used phenomological model of disk black-body emission, \textit{diskbb} \citep{Mitsuda1984} and a model for emission from a warm population of non-thermal electrons (warm corona), namely the thermally comptonized continuum,  \textit{nthComp}  \citep{Zdziarski1996,Zycki1999}.

To treat the ionized, relativistic reflection, we use the \emph{relxillD} model \citep{relxill1,relxill2}, which combines disk reflection with relativistic line blurring in the strong gravitational field of the black hole to account for the blurred reflection near the Fe-K line, and contribute to the soft emission. Crucially, in this model disk density can be varied between  $10^{15}$--$10^{19}$~cm$^{-3}$ \citep{Garcia2016}.  

We tie a number of parameters between these continuum components. Specifically, the incident power-law photon index is tied between the \emph{powerlaw}, \emph{xillver}, and \emph{relxillD} components. Disk iron abundance is tied between \emph{xillver} and \emph{relxillD}. In emission components that require it (\emph{xillver} , \emph{relxillD}), we fix the redshift to the literature value for NGC~4051. A number of \emph{relxillD} parameters are kept at their default values, such as the emissivity for the coronal flavor model (a single power law with $Index=3$, $Rbr=15$ gravitational radii), the inner and outer radii of the accretion disk ($Rin=-1$, meaning ISCO, $Rout=400$ gravitational radii), and the observed cutoff energy ($Ecut=300$~keV), while the black hole spin is restricted to be positive. We only use the reflected component of the \emph{relxillD} model. Additionally, we limit the inclination to be above 20$^\circ$ and below 80$^\circ$, which is strongly supported by jet   \citep[e.g.][]{ Maitra2011} and Narrow Line region \citep{Christopoulou1997} morphologies, and which removes certain model degeneracies. The \emph{xillver} model used is averaged over inclination, black hole spin, and is kept at its default observed cutoff energy of 300~keV.

An additional multiplicative  \textit{constant} is also included in our models and fixed to 1 for MEG spectra, but  kept free for HEG spectra. In this way we account for potential systematic calibration differences in normalization between the two gratings. 

\subsubsection{Absorption by the neutral ISM}

Galactic absorption by the neutral phase of the Interstellar Medium (ISM)  is handled by the \textit{TBabs}\footnote{Note that in the version of XSPEC used, \textit{TBnew} has been renamed to \textit{TBabs}.} model \citep[][]{Wilms2000}.  Potential absorption by ionized phases of the ISM are accounted for by models of absorption by collisionally ionzied gas (cf. following Section). 

\subsubsection{Absorption by collisionally ionized gas}\label{sec:cieabs}

We consider absorption by gas in collisional ionization equilibrium (CIE). This state of gas is uniquely characterized by its temperature $T$. We compute a finely populated grid of absorption models using Cloudy, with temperatures spanning $10^4$--$10^{10}$~K ($\Delta \text{log}~T = 0.1$). This temperature range covers the vast majority of ionic species that can be present in X-ray spectra. 

The hydrogen density is fixed to 1~cm$^{-3}$ in our calculations, since density does not affect collisional ionization balance of ionic species present in X-ray spectra. We vary the hydrogen column density $N_H$ ($\text{log}~N_H/\text{[cm$^{-2}$]}=17.75-23.75$, $\Delta  =  0.25$),  turbulent line width $\sigma$ ($\text{log}~\sigma/\text{[km s$^{-1}$]}=1.3-4.0$, $\Delta =  0.15$), and redshift $z$ (which is equivalent to velocity $v$).

\subsubsection{Absorption by photoionized gas} \label{sec:pie}

We model gas photoionized by the central AGN using Cloudy. As part of our fitting procedure, we utilize an initial pre-computed grid of models, and then compute Cloudy models fully self-consistently (cf.~Section~\ref{sec:mcmc}). The latter required us to marginally modify Cloudy's code to allow calculation of a single absorption table (which is not possible by default)\footnote{See Cloudy users group for details: \url{https://cloudyastrophysics.groups.io/g/Main/topic/57293158}.}.  All models are calculated at a higher resolution (0.3$\times$ the default) to match the resolution of the data.

We use the default Cloudy AGN continuum shape. It consists of a  ``Big Bump'' component at low energies, thought to be disk emission, and an X-ray power-law at high energies. The Big Bump is also a power-law, but is exponentially cut of in the IR, and far UV, where it meets the coronal emission. Photons from this SED component critically affect ionization balance and the resulting models, in particular when collisional processes start to be important \citep[cf.~Section~\ref{sec:disxiden} and ][]{futureHighDen}.

We choose the UV power law slope of 2.0, and the UV cutoff temperature of 10$^5$~K, as informed by observations of this source \cite{Devereux2013}. The relative normalization of these two components, $\alpha_{ox}$, is fixed at  1.17 \cite{Devereux2013}. 

The X-ray power-law photon index, $\Gamma$ is set to 1.9 in the initial grid.(This value is consistent with the HETG data.) In the self-consistent part of our fitting procedure, a value of the power-law index is tied to the spectral indices of the continuum components (coronal power-law and disk reflection). The assumed photoionizing SED is shown in Figure~\ref{fig:SED}.

The photoionization equilibrium (PIE) of the gas is characterized by an ionization parameter $\xi$, as defined by e.g.  \cite{Tarter1969}:
\begin{equation} \label{eq:xi}
    \xi = \frac{L_{ion}}{n_H r^2},
\end{equation}
where $n_H$ is the hydrogen density of the surface of the cloud, $r$ is the distance from the surface to the ionizing source, and $L_{ion}$ is the ionizing luminosity in the 1--1000~Ry range \citep[following e.g.][]{Kallman2001}. 

Further parameters that are varied in the models of photonionized absorption are the gas hydrogen density $n_H$,  hydrogen column density $N_H$, redshift $z$ (which can be translated into wind velocity $v$), and turbulence $\sigma$. Turbulence here is a measure of the non-instrumental width of the line, and it is assumed not to be a source of additional energy or pressure in the gas. The gas density is kept constant as a function of radius in the cloud. We note that the maximum temperature that Cloudy can reach is 10$^{10}$~K.

\begin{table}
\centering
\caption{Parameters of the grid of photoionized models that we use to pre-search the parameter space. All models were calculated with Cloudy ver. 17.0. Note that all values are base-ten logarithms  of the actual quantity, and the spacing is also in log space. \label{tab:grid}}

\begin{tabular}{@{ }c@{  \: }c@{ \: }c@{ \: }c@{ \: }c@{ \: }}
\hline
parameter &  symbol & range  & spacing & unit\\ 
\hline
ionization parameter & $\xi$ & -1 -- 5 &  0.1 & erg cm s$^{-1}$ \\
hydrogen density & $n_H$  & -1--15 & 0.5  & cm$^{-3}$ \\
column density & $N_H$ & 19.2--24 & 0.3 & cm$^{-2}$ \\
turbulence (line width) & $\sigma$ & 1.7--4.5 & 0.155  & km s$^{-1}$ \\
\hline
\end{tabular}
\end{table}

\subsection{Fitting procedure} \label{sec:mcmc}

For each spectral model considered, we identify preferred model parameters using  Markov Chain Monte Carlo (MCMC) methods. Specifically, we use our custom \textsc{Python} fitting procedure, tailored to high performance computer clusters, which uses the \texttt{emcee} package \citep{emcee} for MCMC, the \texttt{schwimmbad} package for handling Message Passing Interface (MPI) parallelization \citep{schwimmbad}, and \texttt{pyXSPEC}, a \textsc{Python} interface to XSPEC, for evaluating the fit statistic.  We use C-statistic \citep{Cash1979} to evaluate model likelihood (modified to include background counts by XSPEC), since a significant fraction of spectral bins have only a few counts and are not in the Gaussian regime.

When considering photoionization models, we employ a two step exploration of the parameter space. Firstly, we pre-search the parameter space with a large, pre-calculated grid of models of photoionized absorption. This grid (summarized in Table~\ref{tab:grid}) covers most of the parameter space that we expect the wind parameters to be in, while the spacing has been chosen based on the data, and is typically a factor of a few smaller than typical error bars on the derived parameters, minimizing the chance that any spectral features are missed. The density starts around the values typical for the interstellar medium, up to the expected  density of accretion disks (from where the wind can possibly be launched), while the ionization parameter spans a range that can produce ions with transitions in the X-ray regime. The column density covers the range detectable as an absorption feature with the current S/N, and reaches the Compton thick regime. The turbulent broadening of the absorption features (which, crucially, is not dissipating energy, but rather is a measure of the line width) starts below the nominal spectral resolution of HETG and goes up to $\sim$30,000 km s$^{-1}$, which is on order of the broadest wind absorption features reported in the literature \citep[e.g.][]{2011ApJ...742...44T}. Note that troughs much broader than few 10,000 km s$^{-1}$ would likely be difficult to distinguish from the global continuum, especially in high energy grating spectra.

Secondly, we use the parameters that are most likely to be the likelihood maximum, as identified using the grid, as starting points of self-consistent MCMC processes. Specifically, at each chain step we vary the continuum parameters, which are then fed into the photoionization calculation. All chains are run to convergence\footnote{In practice, to guarantee convergence of our fits, we need to bound the parameters. This is important in cases where a parameter is unconstrained by the data. We choose conservative parameter limits, which effectively do not restrict the parameter space searched: $\text{log}~\xi=-5-10$, $\text{log}~n_H=-3-20 $, $\text{log}~N_H=15-26 $, and $\text{log}~\sigma= 0.1-5$.}. As a result, we identify the most likely global maximum, its best fit parameters and their posterior distribution, as well as covariances between parameters (including covariances between the continuum emission and wind absorption). Section~\ref{sec:disgrid} summarizes the improvements this step introduces.

\subsection{Bayesian model selection} \label{subsec:dic}

In order to characterize the physical properties of all absorbers that are present in the spectra, we want to quantify how many model components can be robustly constrained. We start by constructing the simplest, physically motivated continuum emission model (Section~\ref{subsec:cont}), and then iteratively add (both collisionally and photo-ionized) wind absorption  components to our global model. As described in the previous section, at each step we find the preferred parameter values with an MCMC procedure and obtain full posterior distributions for all model parameters. 

To evaluate whether an additional spectral component is required by the data, we utilize the Deviance Information Criterion (DIC), as defined by  \cite{Spiegelhalter2002}:
\begin{equation}
\text{DIC} = D(\overline{\theta}) + 2 p_D, \quad p_D = \overline{D(\theta)} - D(\overline{\theta}),
\end{equation}
\noindent where $D(\theta) = -2\text{log}(p(y|\theta))$ is the deviance of a parameter vector $\theta$ given the likelihood of data vector $y$, and $p_D$ is the effective number of model parameters, which is a measure of model complexity. For estimators (barred quantities), we use a mean for deviances, and the multidimensional mode for $\overline{\theta}$, given that some of the final parameter posteriors are lower/upper limits. 

In contrast to other commonly used information criteria, such as Akaike Information Criterion \citep{AIC} and Bayesian Information Criterion \citep{BIC}, DIC uses an effective number of parameters, thus, following the logic of Bayesian evidence, not penalizing for parameters that are unconstrained by the data \citep[for a detailed discussion, see e.g.][]{Liddle2007}. Further, DIC is straightforward to compute from MCMC produced posterior samples (i.e. MCMC chain). 

To establish which model is more likely given the data, i.e. whether an additional spectral component is required, we compare the computed DIC against a scale first proposed by \cite{Jeffreys1961}, and later updated by \cite{Kass1995}. Here, the model with lower DIC is preferred, and the difference in DIC, $\Delta$DIC, is a measure of how strong this preference is. This so-called ``Jeffrey's scale'' is shown in Table~\ref{tab:scale}. 

We add components until the $\Delta$DIC suggests that the next component is no longer statistically significant. We conservatively choose the $\Delta$DIC cutoff of 10, requiring `very strong' evidence to add a model component. 

\begin{table}
\centering
\caption{Adopted interpretation of the difference in DIC, $\Delta$DIC, between two models   \citep[following][]{Kass1995}. In this work we conservatively require $\Delta$DIC of at least 10 to say that one model is preferred over another.  \label{tab:scale}}

\begin{tabular}{@{ }c@{  \: }c@{ \: }}
\hline
$\Delta$DIC&   Evidence against model with higher DIC\\ 
\hline
0 -- 2 &   Not worth more than a bare mention \\
2 -- 6 &  Positive\\
6 -- 10 & Strong \\ 
$>$10 & Very strong \\
\hline
\end{tabular}
\end{table}

\section{Results}\label{sec:results}

\subsection{Best fit global continuum} \label{sec:rescont}

\begin{figure*}
\begin{minipage}{180mm}
\centering
\includegraphics[width=\columnwidth,trim={0mm 0mm 0mm 0mm},clip]{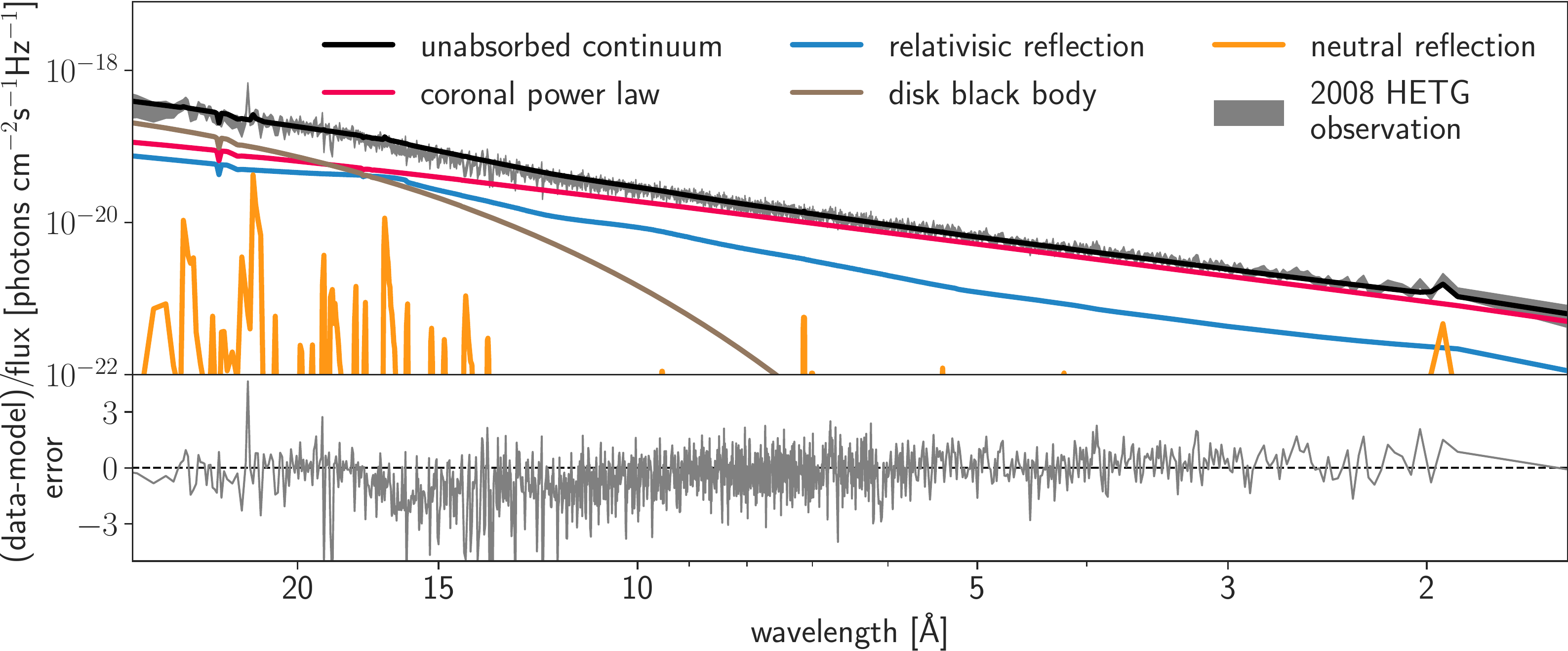}

\vspace{5ex}

\includegraphics[width=\columnwidth,trim={0mm 0mm 0mm 0mm},clip]{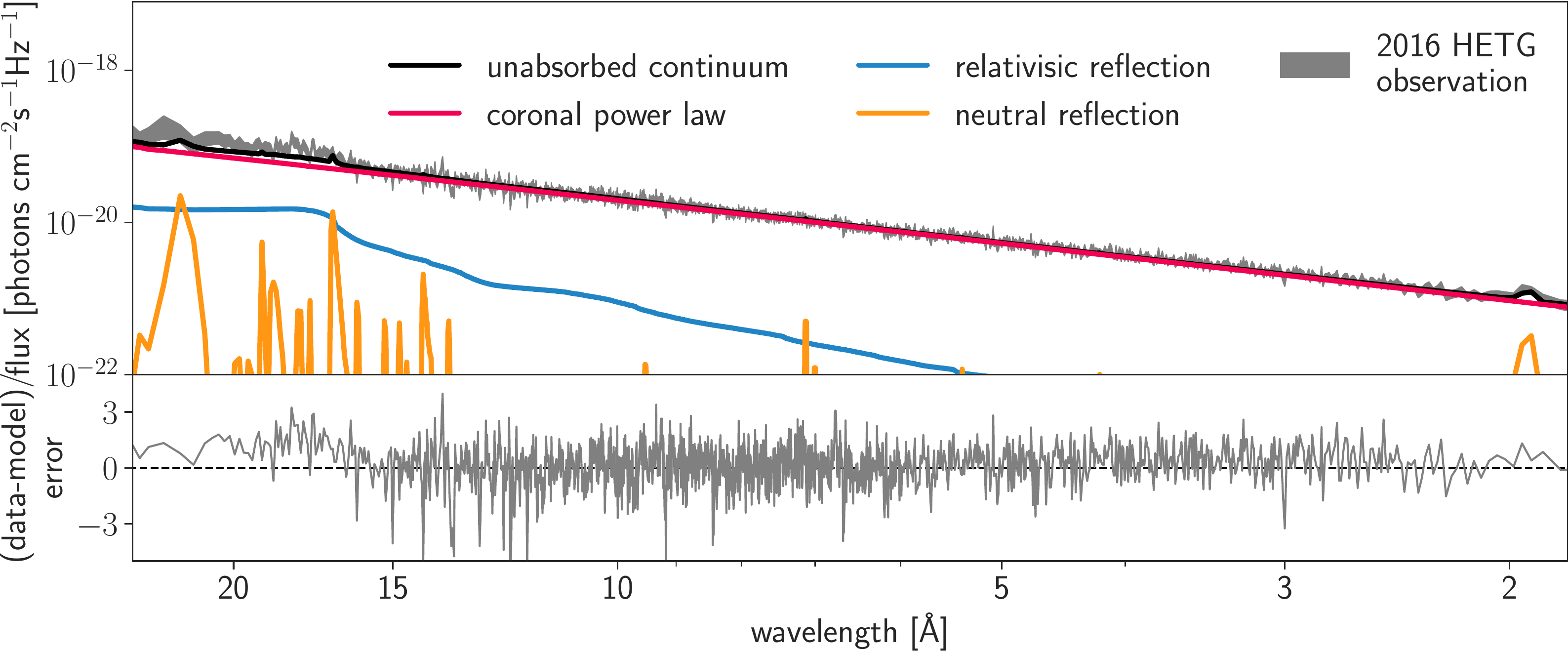}

\caption{Unabsorbed continuum model is shown in black for the 2008 (top) and 2016 (bottom) \textit{Chandra} HETG observation of NGC~4051. Data, grouped for visual purposes, is shown in grey. Spectral components of this model are the coronal power-law (red line), relativistic reflection (blue) line, and neutral reflection (orange). Residuals (data-model/error) are shown in the bottom panels. The soft excess in 2008 data could not be explained with relativistic reflection only, and thus an addition of a phenomonological model, disk black body (brown), was necessary. Continuum model selection and results are discussed in Section~\ref{sec:rescont}, and best fit parameters are shown in Table~\ref{tab:contfit}. Discussion of soft excess modelling is presented in Section~\ref{sec:disrel}. Note that absorption by outflow is not accounted in the model here.  }
\label{fig:cont}
\end{minipage}
\end{figure*}

\begin{table*} 
\begin{minipage}{160mm}
     \centering 
        \caption{$\Delta$DIC values for continuum models tested (with a constant value subtracted for clarity). Note that the tested model also included continuum absorption by neutral ISM (\textit{TBabs}). We show the difference with respect to the model with lowest DIC for both datasets. These values of $\Delta$DIC can be directly compared with Table~\ref{tab:scale} to infer how significant the difference between models is. \label{tab:contDIC} }
 \begin{tabular}{@{}cccccc}  
  \hline
  \textit{powerlaw}+\textit{xillver}+ :  & \textit{diskbb} & \textit{nthComp}  & \textit{relxillD} & \textit{diskbb}+\textit{nthComp} &  \textit{diskbb}+\textit{relxillD}\\
  \hline
    \multicolumn{6}{c}{\textbf{2008 data}} \\ 
\hline 

$\Delta$DIC wrt the best model & -111 & -307&    -92&   -83 &    \textbf{0} \\
\hline
  \multicolumn{6}{c}{\textbf{2016 data}} \\ 
  \hline
$\Delta$DIC wrt the best model & -168 & -309 &  \textbf{0} &    -164 &      -1 \\
\hline
 \end{tabular}
  \end{minipage}
 \end{table*}

We identified the preferred   spectral models for the 2008 and 2016 data sets. Our baseline continuum models are:
\begin{align*}
\text{2008}: &\quad powerlaw+relxillD+diskbb+xillver,  \\
\text{2016}: &\quad powerlaw+relxillD+xillver.
\end{align*}

Using the $\Delta$DIC criterion (Section~\ref{subsec:dic}) we decisively rule out both the \textit{diskbb} (phenomological soft excess) and \textit{nthComp} (warm corona) models as the primary source of the soft excess. This emission component is best described by the high disk density blurred reflection model \textit{relxillD}, with $\Delta$DIC of $\sim$100s over other models. Beyond the DIC criterion, we note that fits to the 2016 data with \textit{relxillD} also result in lower disk iron abundances, as compared to fits with models with low, fixed disk density, warm corona, or phenomenological models \citep[cf.][]{2018MNRAS.477.3711J}. 

We tested for the presence of soft excess emission not accounted for by \textit{relxillD} by adding an additional phenomonological \textit{diskbb} component. The additional component  does not improve the fit for the 2016 epoch ($\Delta$DIC=-1), and the fitted \textit{diskbb} has normalization and temperature consistent with zero, making the component statistically insignificant. The soft excess is stronger in the 2008 spectra, where the additional \textit{diskbb} is clearly statistically required by the data. However, we postulate that the \textit{diskbb} is only required due to the limitations of the \textit{relxillD} model, and with better physical modeling with \textit{relxillD}, the \textit{diskbb} would likely not be required (see Section.~\ref{sec:disrel} for discussion).

\begin{table}

\centering
\caption[Continuum best fit values]{Continuum best fit values with 68\% credible intervals. If a parameter only has a lower/upper limit, then the extreme value with a 95\% credible interval is quoted. \label{tab:contfit} }

\begin{tabular}{@{ }c@{  \: }c@{ \: }c@{ \: }c@{ \: }}
\hline
\hline
\multicolumn{2}{c}{Parameter} & \multicolumn{2}{c}{Best fit value}\\ 
Name & Unit &  2008 & 2016\\
\hline
\hline
\multicolumn{4}{c}{\emph{tbabs}}\\
\hline 

nH & 10$^{20}$~cm$^{-2}$ & 2.0$^{+0.8}_{-0.8}$ & 0.0188$^{+1.6}_{-}$ \\ 
\hline
\multicolumn{4}{c}{\emph{powerlaw}}\\
\hline 
photon index  & - & 1.92$^{+0.02}_{-0.02}$ & 1.89$^{+0.01}_{-0.01}$   \\
normalization & 10$^{-3}$~$\frac{{\rm photons}}{{\rm keV cm}^2 {\rm s}}$ @1keV & 7.6$^{+0.5}_{-0.5}$ & 7.42$^{+0.01}_{-0.01}$ \\
\hline
\multicolumn{4}{c}{\emph{xillver}}\\
\hline 
photon index &   &\multicolumn{2}{c}{tied to \emph{powerlaw}} \\
iron abundance & solar  &3.0$^{+0.8}_{-0.3}$ & 2.8$^{+0.5}_{-0.3}$   \\
ionization & erg cm s$^{-1}$  & 2.3$^{+0.1}_{-0.6}$ & 2.82$^{+0.1}_{-0.4}$  \\
normalization & 10$^{-5}$~$\frac{{\rm photons}}{{\rm keV cm}^2 {\rm s}}$ @1keV  & 2$^{+1}_{-1}$ & 0.6$^{+0.6}_{-0.1}$  \\ 
\hline
\multicolumn{4}{c}{\emph{relxillD}}\\
\hline
photon index &   & \multicolumn{2}{c}{tied to \emph{powerlaw}} \\
iron abundance  &  & \multicolumn{2}{c}{tied to \emph{xillver}} \\
normalization & 10$^{-5}$~$\frac{{\rm photons}}{{\rm keV cm}^2 {\rm s}}$ @1keV &  3.2$^{+0.6}_{-0.5}$ & 0.73$^{+0.09}_{-0.06}$  \\ 
log ionization &  erg cm s$^{-1}$ &3.17$^{+0.70}_{-0.04}$ & 2.69$^{+0.02}_{-0.09}$ \\
black hole spin &  $\frac{cJ}{GM^2}$ &0.88$^{+0.05}_{-0.6}$ & 0.99$^{+0}_{-0.19}$  \\ 
log disk density &  cm$^{-3}$ &18.9$^{+0.1}_{-0.1}$ & 19.0$^{+0}_{-0.1}$   \\
disk inclination &  $^{\circ}$  &55$^{+1}_{-4}$ & 49$^{+1}_{-2}$\\
\hline
\multicolumn{4}{c}{\emph{diskbb}}\\
\hline 

Temperature  & keV &0.137$^{+0.004}_{-0.004}$ &  -  \\
normalization   & $\frac{R_{in}^2 cos\theta}{D_{10}^2}$ & 5310$^{+1360}_{-1000}$ &  -   \\ 
\hline
\multicolumn{4}{c}{\emph{constant}}\\
\hline
factor  & - &  0.951$^{+0.004}_{-0.005}$ & 0.964$^{+0.004}_{-0.003}$  \\ 
\hline
\hline

\end{tabular}
\end{table}

Best fit continuum model components and the data are shown in Figure~\ref{fig:cont}. We summarize the evidence for this continuum model selection in Table~\ref{tab:contDIC}. The best fit parameter values are presented in Table~\ref{tab:contfit}. All continuum parameters are consistent within error bars between the two data sets, except for the disk ionization and normalization of the \textit{relxillD} model, which we discuss in Section~\ref{sec:disrel}.

The most probable value of the neutral hydrogen column density fitted by the \textit{TBabs} model is consistent between the epochs within uncertainties, and consistent with the Galactic HI column density measurement of 1.08$\times$10$^{20}$~cm$^{-2}$ \citep{Kalbera2005}. The most probable value is higher in the 2008 epoch. We did not fix this parameter to the Galactic value, because there may also be additional neutral absorption in NGC~4051. We note that this choice affects the soft part of the spectrum, where the 2008 data requires a phenomenological \textit{diskbb} component.  A clear covariance between the neutral absorption component and the soft emission parameters is observed (\textit{relxillD} normalization, and \textit{diskbb} normalization and temperature). Since soft excess is stronger in the 2008 epoch, these covariances also explain the higher neutral hydrogen column density preferred by the model.

\begin{figure*}
\begin{minipage}{180mm}
\centering
\includegraphics[width=0.95\columnwidth,trim={0mm 0mm 0mm 0mm},clip]{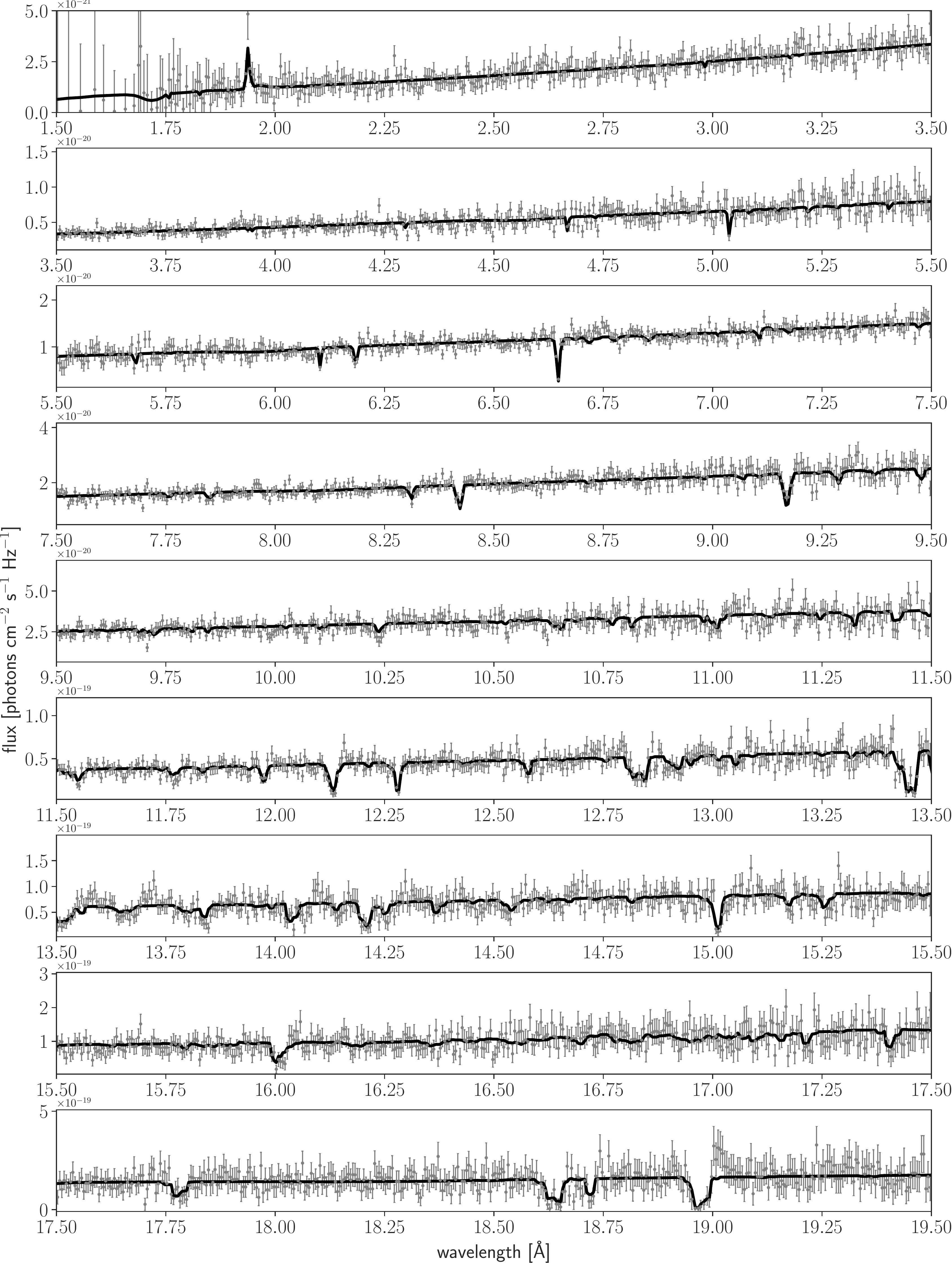}
\caption{ The 2008 HETG/MEG spectrum of NGC~4051 (grey) with the best fit model (black). Absorption features from individual absorbers, along with their ionic species, are shown in Appendix~\ref{sec:absfigs}. }
\label{fig:spec08}
\end{minipage}
\end{figure*}

\begin{figure*}
\begin{minipage}{180mm}
\centering
\includegraphics[width=0.95\columnwidth,trim={0mm 0mm 0mm 0mm},clip]{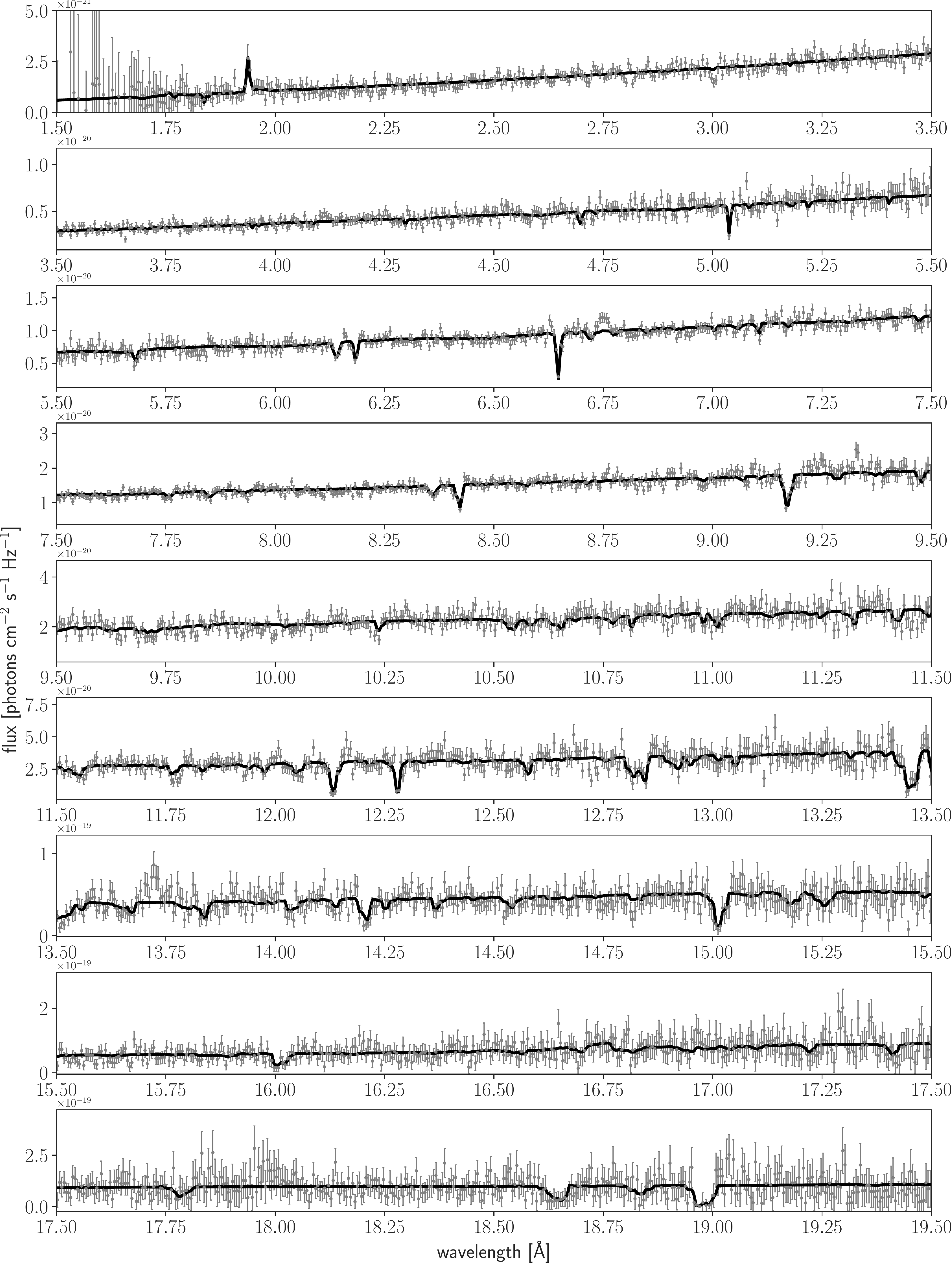}
\caption{ The 2016 HETG/MEG spectrum of NGC~4051 (grey) with the best fit model (black). Absorption features from individual absorbers, along with their ionic species, are shown in Appendix~\ref{sec:absfigs}.  }
\label{fig:spec16}
\end{minipage}
\end{figure*}

\begin{table*} 
\begin{minipage}{160mm}
     \centering 
    \caption[Properties of the absorbers detected in NGC~4051]{Properties of the absorbers detected in NGC~4051. All uncertainties are 68\% credible intervals. We note that for FAST1, the only component that is found to be purely in CIE, the PIE model was statistically not preferred, with $\Delta$DIC$_{2008}$=189 and  $\Delta$DIC$_{2016}$=203 in favour of CIE (cf. Table~\ref{tab:scale}). The choice of CIE for this component is further discussed in Section~\ref{sec:coldis}. \label{tab:outparams}}
 \begin{tabular}{@{}ccccccccc}  
  \hline

identifier & $\Delta$DIC$^{a}$ &  ionization &  velocity & log ionization & log temp. & log density &log col. density & log line width  \\ 
& & equilibrium & /[km s$^{-1}$] & /[erg cm s$^{-1}$]& /[K]& /[cm$^{-3}$]& /[cm$^{-2}$] & /[km s$^{-1}$]  \\
 \hline

    \multicolumn{9}{c}{\textbf{2008 data}} \\ 
\hline \\
SLOW & 43 & CIE & 420$^{+40}_{-40}$ &  - & 5.58$^{+0.03}_{-0.03}$ &  - & 20.6$^{+0.1}_{-0.1}$ & 1.9$^{+0.2}_{-0.2}$    \\
 & 37 & PIE & 480$^{+40}_{-50}$ & 0.51$^{+0.09}_{-0.14}$ &  - & 3$^{+4}_{-4}$ & 20.5$^{+0.1}_{-0.1}$ & 2.0$^{+0.1}_{-0.2}$    \\\\

FAST1 & 653 & CIE & 890$^{+10}_{-10}$ &  - & 6.99$^{+0.02}_{-0.01}$ &  - & 21.4$^{+0.04}_{-0.03}$ & 2.04$^{+0.03}_{-0.1}$  \\\\

FAST2 & 31 & PIE &970$^{+40}_{-30}$ & 1.59$^{+0.06}_{-0.06}$ &  - & 5$^{+2}_{-4}$ & 20.6$^{+0.1}_{-0.1}$ & 2.0$^{+0.1}_{-0.9}$  \\\\

FAST3 & 35 & CIE & 5000$^{+300}_{-100}$ &  - & 7.21$^{+0.04}_{-0.05}$ &  - & 21.4$^{+0.1}_{-0.2}$ & 3.1$^{+0.2}_{-0.3}$  \\
& 32 & PIE & 5000$^{+300}_{-200}$ & 3.0$^{+0.2}_{-0.1}$ &  - & -0.9$^{+4}_{-0.01}$ & 21.5$^{+0.3}_{-0.4}$ & 3.1$^{+0.3}_{-0.3}$    \\\\

VFAST1 & 25 & PIE & 10000$^{+2000}_{-1000}$ & 4.99$^{+0.02}_{-0.26}$ &  - & 2$^{+1}_{-2}$ & 23.8$^{+0.1}_{-0.2}$ & 3.8$^{+0.1}_{-0.1}$ \\\\

\hline
  \multicolumn{9}{c}{\textbf{2016 data}} \\ 
  \hline \\
SLOW & 25  & PIE & 500$^{+60}_{-60}$ & 0.6$^{+0.1}_{-0.2}$ &  - & 6$^{+4}_{-3}$ & 20.5$^{+0.1}_{-0.1}$ & 2.0$^{+0.2}_{-0.2}$  \\\\

FAST1 & 675 & CIE & 870$^{+20}_{-20}$ &  - & 7.01$^{+0.01}_{-0.01}$ &  - & 21.4$^{+0.1}_{-0.04}$ & 2.0$^{+0.1}_{-0.1}$     \\\\

FAST2 & 12 & PIE &930$^{+60}_{-60}$ & 1.7$^{+0.2}_{-0.1}$ &  - & 5$^{+3}_{-3}$ & 20.4$^{+0.2}_{-0.2}$ & 2.0$^{+0.2}_{-0.1}$   \\\\

FAST3 & 92 & PIE & 2960$^{+40}_{-60}$ & 3.2$^{+0.2}_{-0.1}$ &  - & 9$^{+1}_{-2}$ & 22.0$^{+0.2}_{-0.2}$ & 2.6$^{+0.2}_{-0.2}$   \\\\

VFAST1 & 37 & PIE & 9700$^{+400}_{-600}$ & 1.52$^{+0.03}_{-0.02}$ &  - & 13.0$^{+0.01}_{-0.02}$ & 22.8$^{+0.1}_{-0.1}$ & 3.7$^{+0.1}_{-0.2}$    \\
& 35 & CIE & 9600$^{+500}_{-800}$ &  - & 7.18$^{+0.02}_{-0.02}$ &  - & 21.6$^{+0.1}_{-0.1}$ & 3.56$^{+0.1}_{-0.1}$  \\\\

VFAST2 & 37 & CIE & 27400$^{+500}_{-600}$ &  - & 7.4$^{+0.1}_{-0.1}$ &  - & 21.7$^{+0.1}_{-0.1}$ & 3.4$^{+0.2}_{-0.1}$     \\
& 37 & PIE & 27000$^{+1000}_{-2000}$ & 3.1$^{+0.2}_{-0.1}$ &  - & 4$^{+3}_{-3}$ & 22.0$^{+0.4}_{-0.1}$ & 3.8$^{+0.2}_{-0.2}$    \\

\hline
 \end{tabular}
 \footnotesize{(a) $\Delta$DIC is the difference between the DIC of the best fitting model without the  the component and with the addition of the component (cf. Section~\ref{subsec:dic}. The  $\Delta$DIC value for FAST1 component is large, because it is the first absorber identified by our analysis, so here $\Delta$DIC=$\Delta$DIC$_{continuum}$-$\Delta$DIC$_{FAST1*conitnuum}$. This large value means that there is clearly a lot of absorption features in the data. It does not mean that most of the absorption features come from just this one component.  } 
  \end{minipage}
 \end{table*}

\subsection{Outflow structure}
\label{sec:resout}
We robustly detect five absorbers intrinsic to NGC~4051 in the 2008 and six in the 2016 spectra. In addition, in both epochs we robustly identify the second most significant absorption to be the  Milky Way hot halo component, which is a subject of a separate publication \citep[][]{futureMW}. 

We show the full spectra from both epochs in Figures~\ref{fig:spec08} and~\ref{fig:spec16} with best fit model overlaid. Many absorption features are clearly visible, as well as some emission lines that we do not study in the current work. There are no significant features  missed by our modelling, which is what we expect having exhausted the DIC criterion (i.e. no additional components were required by the data). All individual absorbers from both epochs are further shown in Appendix~\ref{sec:absfigs}, along with lists of most significant features in each component and their ionic species. 

We have determined that one absorption component in both the 2008 and 2016 data is purely collisionally ionized (see Section~\ref{sec:coldis}), while two in 2008 and three in 2016 are photoionized. For the remaining components, a PIE and CIE models are equally likely given our adopted criteria (cf.~Section~\ref{subsec:dic} and Table~\ref{tab:scale}). The evidence for each component, $\Delta$DIC, are listed in Table~\ref{tab:outparams}. Please note that $\Delta$DIC is the difference between the best fitting model with and without the particular absorber. This means that its value will be largest for the most statistically significant absorber (here, FAST1), since it will be the difference between pure continuum model and continuum with one absorber. For the next most statistically significant wind component, $\Delta$DIC will be the difference between model with one and two absorbers, and so on. 

\begin{figure*}
\begin{minipage}{180mm}
\centering
\includegraphics[width=\columnwidth,trim={0mm 0mm 0mm 0mm},clip]{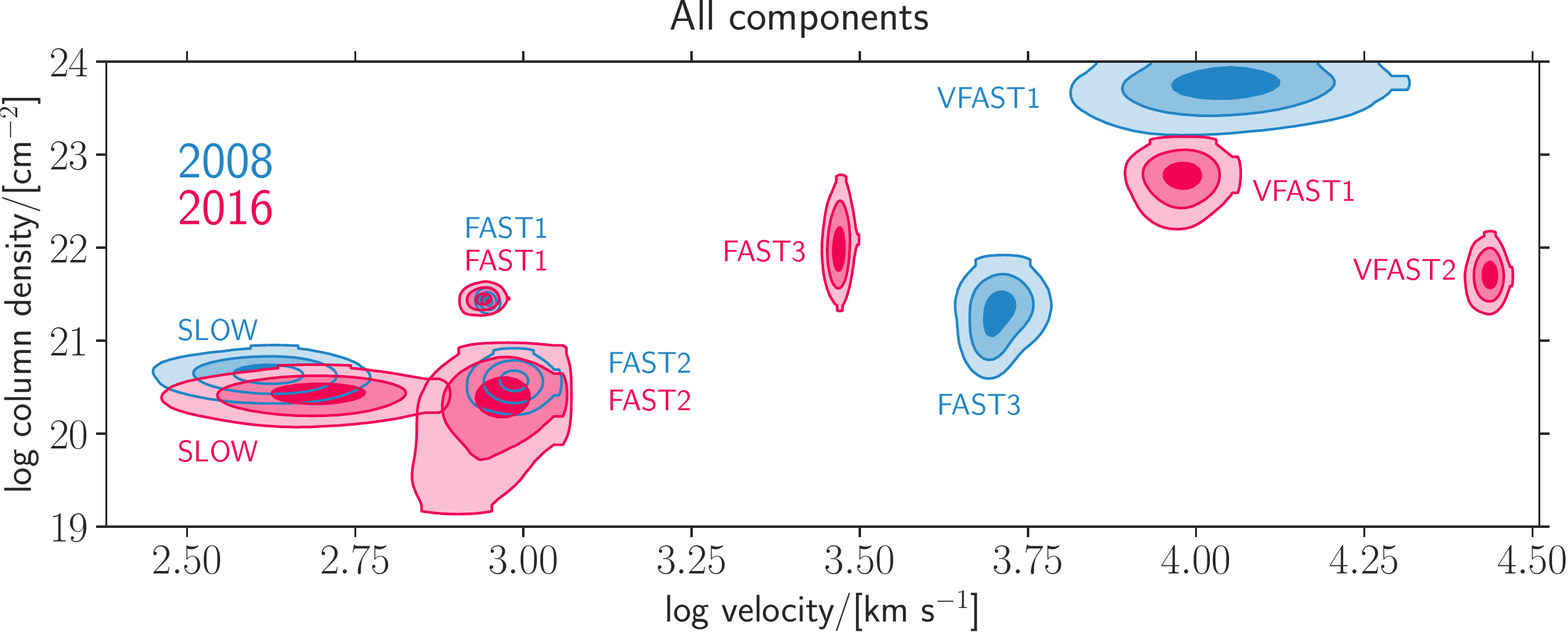}

\vspace{5ex}

\includegraphics[width=\columnwidth,trim={0mm 0mm 0mm 0mm},clip]{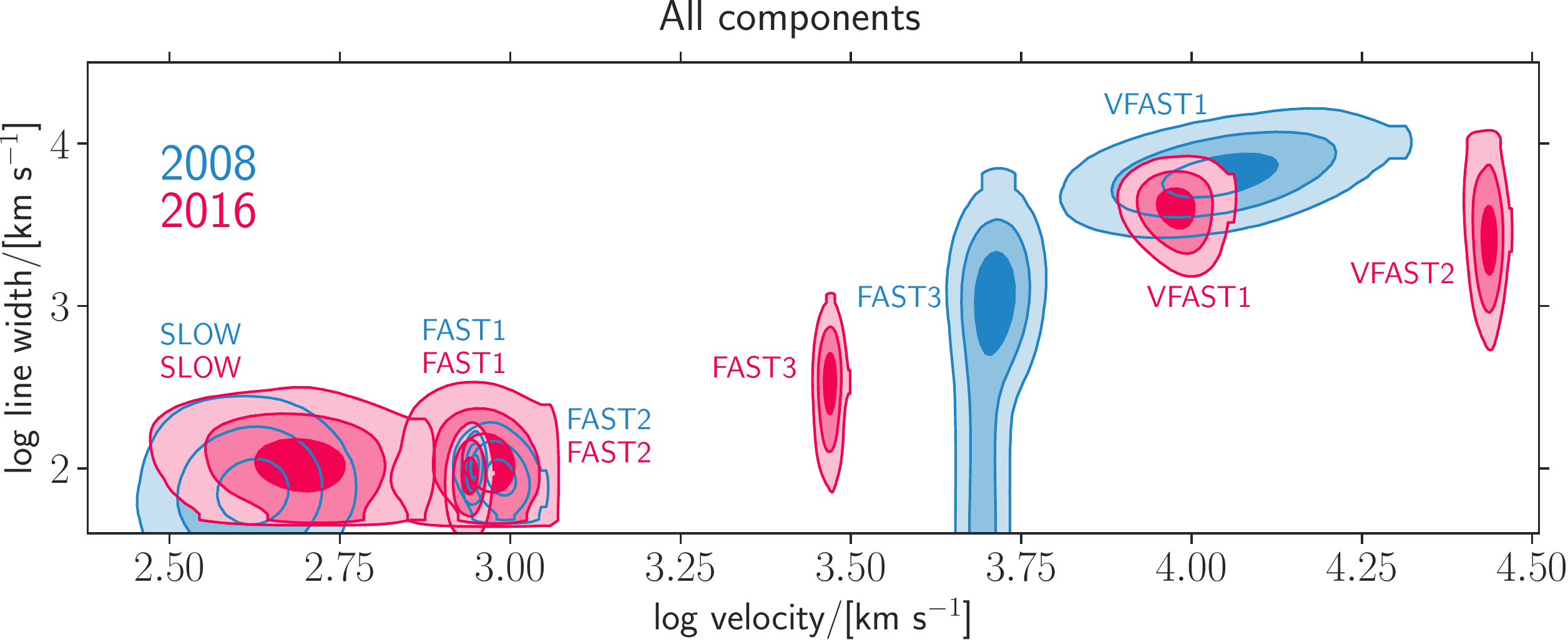}

\caption[2D best fit posteriors for all components]{2D posterior distributions on column density (top), line width (bottom) and velocity. Contours are 1, 2, and 3$\sigma$. Components found in the 2008 and 2016 spectra are shown in blue and red respectively. Remarkably, components with velocities of less than 1000~km~s$^{-1}$ remain consistent between two epochs. }
\label{fig:2Dall}
\end{minipage}
\end{figure*}

\begin{figure*}
\begin{minipage}{180mm}
\centering
\includegraphics[width=\columnwidth,trim={0mm 0mm 0mm 0mm},clip]{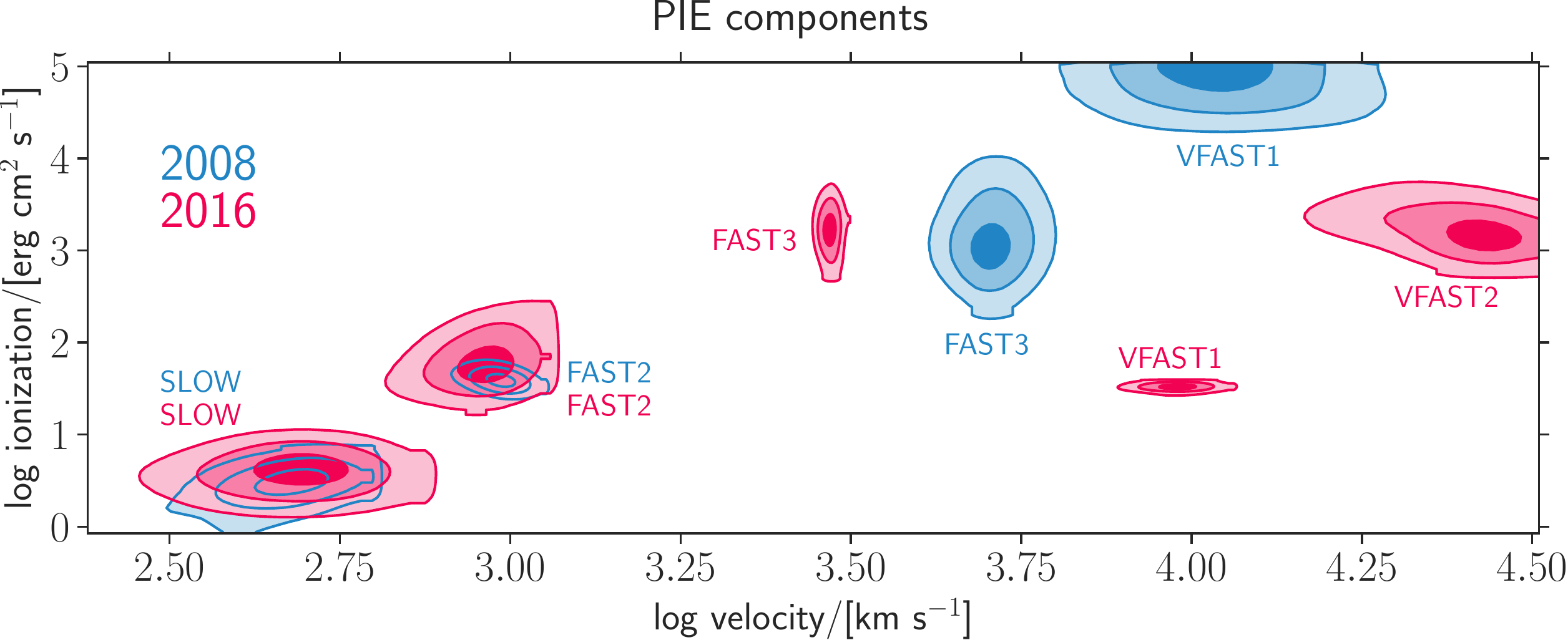}
\includegraphics[width=0.66\columnwidth]{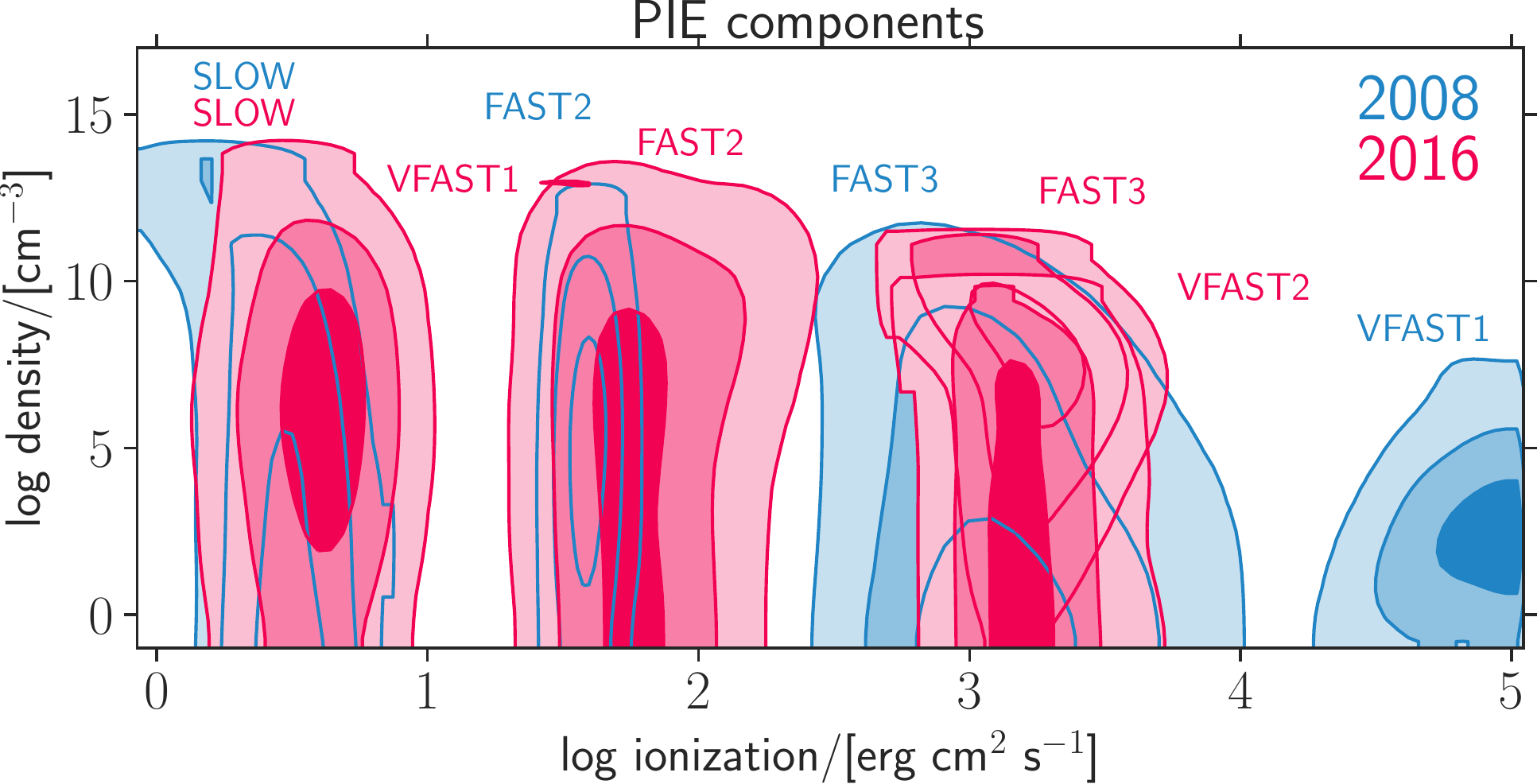}
\includegraphics[width=0.29\columnwidth]{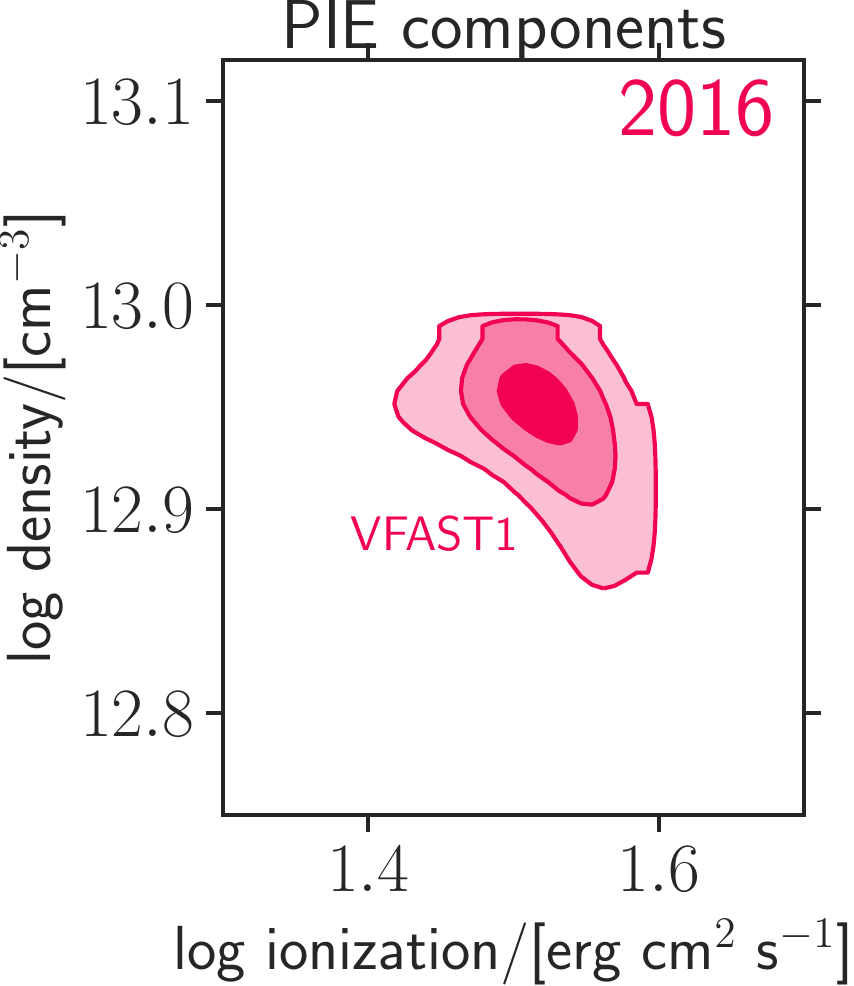}
\caption{\textit{Top}: 2D posterior distributions on ionization and velocity for the photoionized components detected. Contours are 1, 2, and 3$\sigma$. Components found in the 2008 and 2016 spectra are shown in blue and red respectively.  \textit{Bottom}: 2D posterior distributions on density and ionization for the photoionized components detected. Contours are 1, 2, and 3$\sigma$. Components found in the 2008 and 2016 spectra are shown in blue and red respectively. Left panels shows all PIE absorbers, and right panels zooms in onto the 2016 VFAST1 absorber for which we obtain the tightest constraints. }
\label{fig:2DPIE}
\label{fig:2DPIEdenxi}
\end{minipage}
\end{figure*}

\begin{figure*}
\begin{minipage}{180mm}
\centering
\includegraphics[width=\columnwidth,trim={0mm 0mm 0mm 0mm},clip]{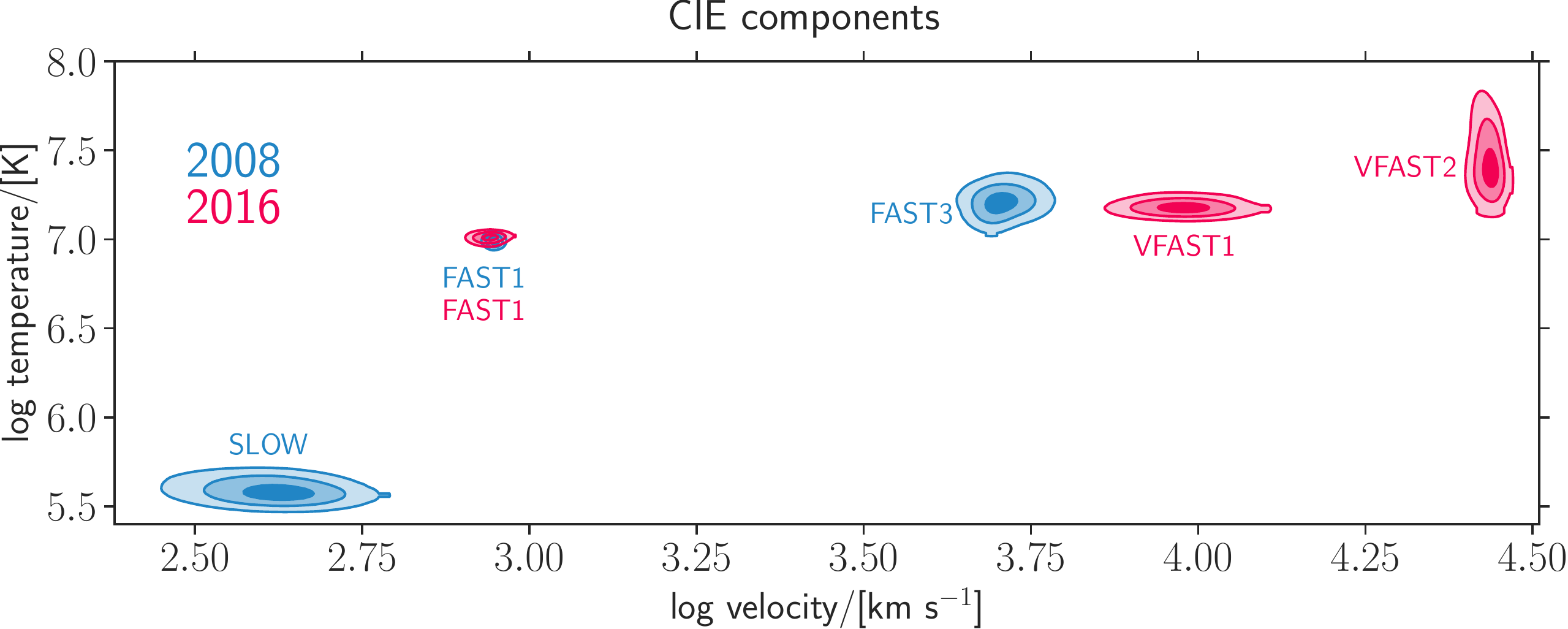}

\caption[2D best fit posteriors for all components]{2D posterior distributions on temperature and velocity for the collisionally ionized components detected. We note that only FAST1 component in both epochs is unambiguously in collisional ionization balance, while for other components shown here, both CIE and PIE models are equally likely (cf.~Table~\ref{tab:outparams}). Contours are 1, 2, and 3$\sigma$. Components found in the 2008 and 2016 spectra are shown in blue and red respectively. }
\label{fig:2DCIE}
\end{minipage}
\end{figure*}

The derived physical properties of outflow components are shown in Table~\ref{tab:outparams}, and in Figures~\ref{fig:2Dall}, \ref{fig:2DPIE}, and~\ref{fig:2DCIE}. Note that we apply relativistic correction when calculating velocities from the directly measured redshifts. 

For cases where we could not discriminate between PIE and CIE, we list both solutions. Note that in these cases the absorbers are very similar in terms of their velocity and column density (for further discussion see Section~\ref{sec:disxiden}).

Most absorbers are clearly and independently present in both epochs, or have similar counterparts, which we determine based on measured velocities and line widths. We name the absorbers according to their velocity, and use the same names for both epochs (cf.~Table~\ref{tab:outparams}). This does not necessarily mean that we are observing the same exact gas in both observations, which we discuss in Section~\ref{sec:distemporal}. 

The one component without a counterpart is the fastest absorber in the 2016 data (VFAST2). Interestingly, there is a tentative matching absorber in the 2008 data, but given our conservative approach, the evidence was not sufficient for us to include it in our best fit model ($\Delta$DIC=6). 

We have also detected a possible highly relativistic component in both spectra, with velocities spanning $\sim$1/3-2/3$c$ (where $c$ is the speed of light). However, in both datasets the feature driving the detection was a single broad absorption trough around 15-18~\AA, which is where the relativistic disk reflection and disk black body models transition from being higher to lower than the coronal power law (cf.~Figure~\ref{fig:cont}). In this particular region of the spectrum where two continuum components meet, if an artificial broad absorption feature was introduced, the fit would easily increase continuum emission to compensate for such broad absorption, masking the fact that it's not physical. Because soft excess is the least physically motivated part of our spectrum, and because this potential absorber appeared in a region where an artificial trough could appear due to the continuum components we chose, we do not claim to have detected this absorber. Confirmation of such a detection will require better understanding of soft excess and higher throughput X-ray spectrometers.

Below we summarize the best fit physical parameters of the detected wind components in more detail. If a CIE and PIE models are equally well fitting a particular component, we discuss their joint properties (as the velocity, line width and column density are consistent between the CIE and PIE solutions).

\subsubsection{SLOW absorber }

We find one slow absorber in both datasets, with velocity $v\approx 450$~km~s$^{-1}$  consistent across both epochs. It consists of shallow, narrow absorption features (log $N_H$/[cm$^{-2}$]$\approx$20.5; $\sigma\approx 100$~km~s$^{-1}$). This absorber is clearly photoionized in 2016, but both CIE and PIE are equally likely in 2008. It is remarkably stable across two epochs. Both ionization equilibria models measure consistent velocity, column density and line widths. This slow wind appears to be relatively cool, at log~$T/$[K]=$5.58^{+0.03}_{-0.03}$ and log $\xi/$[erg cm s$^{-1}$]$\approx$0.5-0.6.

\subsubsection{FAST absorbers }

There are three absorbers present in the data with velocities of the order of 1000~km s$^{-1}$. Two of them, FAST1 and FAST2, are fully consistent between both epochs. Their velocities are  $\sim$880~km s$^{-1}$ and $\sim$950~km s$^{-1}$, respectively. While close in speed, both components are clearly distinguished by different column densities, which differ by over an order of magnitude (cf.~Figure~\ref{fig:2Dall}, top).

FAST1 is the most statistically significant absorber in the spectrum and we decisively determine it to be collisionally ionized with $\Delta$DIC=189 and  $\Delta$DIC=203 for CIE over PIE in 2008 and 2016 respectively (cf. Table~\ref{tab:scale}; for discussion of the origin of this component, see Section~\ref{sec:coldis}). In both epochs features from this absorber are deep, with log $N_H$/[cm$^{-2}$]$\approx$21.4, and narrow, with $\sigma\approx 100$~km~s$^{-1}$. It is also hot, with temperature log $T/$[K]$\approx$7 (significantly higher than the SLOW component). In contrast, FAST2 is in PIE. This component is also narrow ($\sigma\approx 100$~km~s$^{-1}$), but the features are shallower than for FAST1, log $N_H$/[cm$^{-2}$]$\approx$20.5. The ionization parameter is log $\xi/$[erg cm s$^{-1}$]$\approx$1.6, and the density is only weakly constrained (log $n/$[cm$^{-3}$]$\approx$1--8).

Component FAST3 shows variability, changing its velocity from  $5000^{+300}_{-100}$~km~s$^{-1}$ in 2008 to $2960^{+40}_{-60}$~km~s$^{-1}$ in 2016. It is in full PIE in 2016, but both CIE and PIE models provide equally good fit to the 2008 data. Its ionization remains the same at log $\xi/$[erg cm s$^{-1}$]$\approx$3.1. The most probable density increased  between 2008 and 2016, from log $n/$[cm$^{-3}$]=$-0.9^{+4}_{-0.01}$ to log~$n/$[cm$^{-3}$]=$9^{+1}_{-2}$ (but is still consistent at the 2$\sigma$ level). Temperature of the CIE model is log~$T/$[K]=7.21$^{+0.04}_{-0.05}$, slightly hotter than the FAST1 component. The most probable column density increased from log~$N_H$/[cm$^{-2}$]$\approx$21.5 in 2008 to log~$N_H$/[cm$^{-2}$]=22$^{+0.2}_{-0.2}$ in 2016 (but is consistent on 2$\sigma$ level), while line width decreases from $\sigma\approx 1000$~km~s$^{-1}$ to $\sigma\approx 400$~km~s$^{-1}$. 

Note that we may be observing different gas in both epochs, as it moves in and out of the line of sight, rather than the same component evolving over time.

\subsubsection{VFAST absorbers}

We detect two absorbers with velocities that are relativistic (only one in the 2008 data). VFAST1 component has velocity consistent between both epochs, $\sim 10,000$~km~s$^{-1}$ ($\sim$3\%$c$). It is in PIE in 2008, but in 2016 both PIE and CIE models are equally likely. Measured CIE temperature is log~$T/$[K]=7.18$^{+0.02}_{-0.02}$, similar to the FAST3 2008 measurement, and slightly above FAST1 component. Ionization decreases, dropping from log~$\xi/$[erg cm s$^{-1}$]=4.99$^{+0.02}_{-0.26}$ in 2008 to log~$\xi/$[erg cm s$^{-1}$]=$1.52^{+0.03}_{-0.02}$ in 2016, accompanied by a significant change in density, which increases from log~$n/$[cm$^{-3}$]=$2^{+1}_{-2}$ to log~$n/$[cm$^{-3}$]=13.0$^{+0.01}_{-0.02}$. While the 2016 density constraint is precise, it is still consistent at the 3$\sigma$ level with 2008 measurement, which has large error bars.  Column density and line width also change, from log~$N_H$/[cm$^{-2}$]=$23.9^{+0.1}_{-0.2}$ to log~$N_H$/[cm$^{-2}$]$\approx$22, and $\sigma\approx 1200$~km~s$^{-1}$ to $\sigma\approx 4000$~km~s$^{-1}$. We emphasize that the current data cannot distinguish between scenarios where the same absorber evolves in time and where we are observing two unconnected absorbers moving in and out of our line of sight.

The fastest  detected absorber in our data, VFAST2, is only robustly present in the 2016 epoch. There are hints of its counterpart in the 2008 spectra, but this dataset did not have enough counts to confirm a detection. We determine its velocity to be $\sim 27,200$~km~s$^{-1}$ ($\sim$9\%$c$). There is not enough signal in the data to distinguish between PIE and CIE for this component. Its temperature log~$T/$[K]=$7.4^{+0.1}_{-0.1}$, is slightly higher than VFAST1 and FAST3. It is highly ionized, log~$\xi/$[erg cm s$^{-1}$]=$3.1^{+0.2}_{-0.1}$, and not dense, log~$n/$[cm$^{-3}$]=$4^{+3}_{-3}$. It has high column density, log~$N_H$/[cm$^{-2}$]$\approx$21.8 and is the broadest of all absorbers, with $\sigma\approx 4000$~km~s$^{-1}$.

\section{Systematic uncertainties}\label{sec:uncertainties}

\subsection{Atomic data and choice of models}
\label{sec:atomic}
There are three major PIE models used currently to study X-ray absorption by AGN outflows: Cloudy \citep{Ferland2017}, XSTAR \citep{Kallman2001} and \texttt{pion}/SPEX \citep{Mehdipour2016}. In this work, we used Cloudy since it offers most control over physical set up of the model, which enabled self-consistent modelling of the AGN spectrum, and because it was easy to control  from within our \textsc{Python} based inference framework \citep[note that the recent SPEX version, 3.06.00, also offers a \textsc{Python} interface, ][]{kaastra_j_s_2020_3939056}. We further use Cloudy to model hot CIE absorption, as this was the most up to date model available via XSPEC fitting package. 

\cite{Mehdipour2016} have done a systematic comparison of these three photoionization codes and found that there is up to 30\% difference in the best fit ionization parameter. This systematic uncertainty should be added to ionization parameters measured in our work. 

\cite{2018PASJ...70...12H} used the first Hitomi X-ray microcalorimeter spectrum of Perseus Cluster, which had unprecedented resolution and S/N, to compare the completeness of databases underlying these codes, and found that Cloudy was lacking a number of X-ray lines in its databases. Since the absorbers we detected are mostly driven by well-known transitions, and our spectrum is not as high S/N as the Perseus spectrum, this should have small impact on our measurements. However, it will be crucial to use models based on more complete atomic data when analyzing spectra from future X-ray missions. 

\subsection{Mixed ionization gas} \label{sec:disxiden}

For high density gas (above $n\approx10^{7-10}$~cm~$^{-3}$, depending on the particular SED), collisional processes will start to be important \citep[see also][]{2021ApJ...908...94K,futureHighDen}. This means that there is a transition region between pure PIE and pure CIE plasma, where both processes are important. In this case, it is possible to determine the density of the gas precisely, as for the VFAST1/2016 absorber \citep[][]{futureHighDen}.

In principle, because Cloudy is a fully self-consistent spectral synthesis code, it should  model this transition region well. However, as discussed in previous Section, the code may be missing some atomic transitions, and the data may not have enough constraining power to constrain this narrow part of parameter space. As a result, for a gas where collisional processes are starting to be as important as photoionization, both PIE and CIE models could appear to fit the data well. 

Additionally, modest S/N spectra or the lack of significant distinguishing features can also make it hard to distinguish between PIE and CIE models.

\subsection{Assumed SED}
\label{sec:unc_SED}
We have based the choice of photoionizing SED on available multiwavelength observations of NGC~4051 (cf. Section~\ref{sec:pie}). However, it is not necessary that each absorber is exposed to the same SED, as especially the lower energy photons may be emitted from regions that are further from the black hole than the particular wind component. 

To estimate if this would significantly change our results, we checked if the PIE measurements would change if the low energy part of our SED was lowered by a few orders of magnitude. We found that this results in slightly lower upper estimates on density, and that this difference is negligible compared to the already large uncertainties on this parameter.

\section{Discussion}\label{sec:discussion}

\subsection{Comparison with previous works}

NGC~4051 has been observed with both \textit{Chandra} and XMM-\textit{Newton} X-ray gratings in a number of works. M of these analyses show similar absorbers to our SLOW and FAST components. While the velocities are similar, the derived column densities and ionization parameters vary. This is likely because many previous works assumed simplified SEDs, and/or fixed fit parameters, especially density and line width, which we leave free to vary. Additionally, we expect that different spectral codes will differ by up to 30\% in ionization parameter \citep[cf.~Section~\ref{sec:atomic} and ][]{Mehdipour2016}. Note also that none of the analyses consider that the gas could be in collisional ionization equilibrium, thus implicitly assuming that the radiation field is the dominant source of ionization.

\cite{King2012} have analyzed the 2008 \textit{Chandra} HETG used in this work. They fixed line width at 200~km~s$^{-1}$, thus their analysis was sensitive to narrow absorbers only. They found an absorber with $v=400^{+270}_{-380}$~km~s$^{-1}$ which is consistent with SLOW/2008 component, with ionization log($\xi$/[erg cm s$^{-1}$])] = 1$^{+0.06}_{-0.15}$, which is higher than measured in our analysis. This discrepancy can be attributed to the fact that in their analysis density was fixed at $10^{10}$~cm$^{-3}$, while we left density free to vary and found it to be lower. They also find two components with velocities $1000^{+30}_{-70}$~km~s$^{-1}$ and $1090^{+120}_{-110}$~km~s$^{-1}$. The measured column densities suggest that these absorbers are the FAST1/2008 and FAST2/2008 components. Further, \cite{King2012} report on three additional absorbers with velocities 600--700~km~s$^{-1}$, which we attribute to absorption by the hot halo of the Milky Way (cf.~Section~\ref{sec:dismw}).

\cite{Lobban2011} also analyzed the 2008 \textit{Chandra} HETG dataset. They considered line widths of  200~km~s$^{-1}$ and 500~km~s$^{-1}$. They find two absorbers with $v=180\pm$100~km~s$^{-1}$,  $\sigma=200$~km~s$^{-1}$ and $v=220^{+60}_{-40}$~km~s$^{-1}$ with $\sigma=500$~km~s$^{-1}$ which we do not find in our work. These components could be based on features of our SLOW/2008 component. They also report on an absorber with $v=550\pm$60~km~s$^{-1}$,  $\sigma=200$~km~s$^{-1}$ which could be a mix of the SLOW/2008 component and the Milky Way hot halo absorption. They further report on two absorbers with $v=820\pm$30~km~s$^{-1}$,  $\sigma=200$~km~s$^{-1}$ and $v=710^{+40}_{-20}$~km~s$^{-1}$,  $\sigma=200$~km~s$^{-1}$, which are most likely our FAST1/2008 and FAST2/2008 absorbers. They also find a broad absorber with fixed line width of 3000~km~s$^{-1}$ and velocity $5800^{+1200}_{-860}$~km~s$^{-1}$. This is consistent with our FAST3/2008 absorber. However we find this absorber to be narrower, with lower column density and ionization parameter.  These differences between the two analyses can also be explained by fixing line width and density. We also note that \cite{Lobban2011} analysis assumed a simple power law with $\Gamma$=2.5, which is much higher than what our analysis and \cite{King2012} have found,  $\Gamma$ of about 1.9. This could further contribute to difference in measured absorbers.

Analyses of data from different instruments and epochs also find the slow and narrow absorbers that we find in our work \citep[e.g.][]{Krongold2007, Steenbrugge2009, Pounds2011, Silva2016, Pounds2013}. This is not surprising, since our analysis shows that these outflow components remain stable between the two epochs analyzed in this work.

Fast components have been found in a number of work. As mentioned above, \cite{Lobban2011} detect our FAST3/2008 absorber in the same data as analyzed here. Analyzing a shorter HETG exposure from 2000, \cite{Collinge2001} detect absorber moving at 2340$\pm$130~km~s$^{-1}$. \cite{Silva2016} find two absorbers similar to FAST3/2008 with velocities 4260$\pm$60~km~s$^{-1}$ and 5770$\pm$30~km~s$^{-1}$ in the 2009 XMM-\textit{Newton} RGS spectra. \cite{Pounds2013} find absorbers with $v=5760\pm500$~km~s$^{-1}$ and $v=3720\pm300$~km~s$^{-1}$ in the same data. Also using the 2009 RGS spectra, \cite{Mizumoto2017} report wind components with $v=4060\pm60$~km~s$^{-1}$ and $v=6120\pm20$~km~s$^{-1}$ . All of these analyses make different assumptions about the SED and fix a number of parameters (most commonly line width and density). 

Very fast absorber is detected in the literature only by \cite{Pounds2013}. In the 2009 XMM-\textit{Newton} RGS spectra they find a component moving at $v=10290\pm1000$~km~s$^{-1}$, which is consistent with our VFAST1/2008 component at better than 1$\sigma$. However, their measured ionization parameter and column density differ by two orders of magnitude from ours. This is likely due to fixing fit parameters as well as the fact that RGS spectra don't go beyond 2~keV, while HETG extends beyond 8~keV with good statistics.

\subsection{Collisionally ionized absorbers}
\label{sec:coldis}

\begin{figure*}
\begin{minipage}{180mm}
\centering
\includegraphics[width=\columnwidth,trim={0mm 0mm 0mm 0mm},clip]{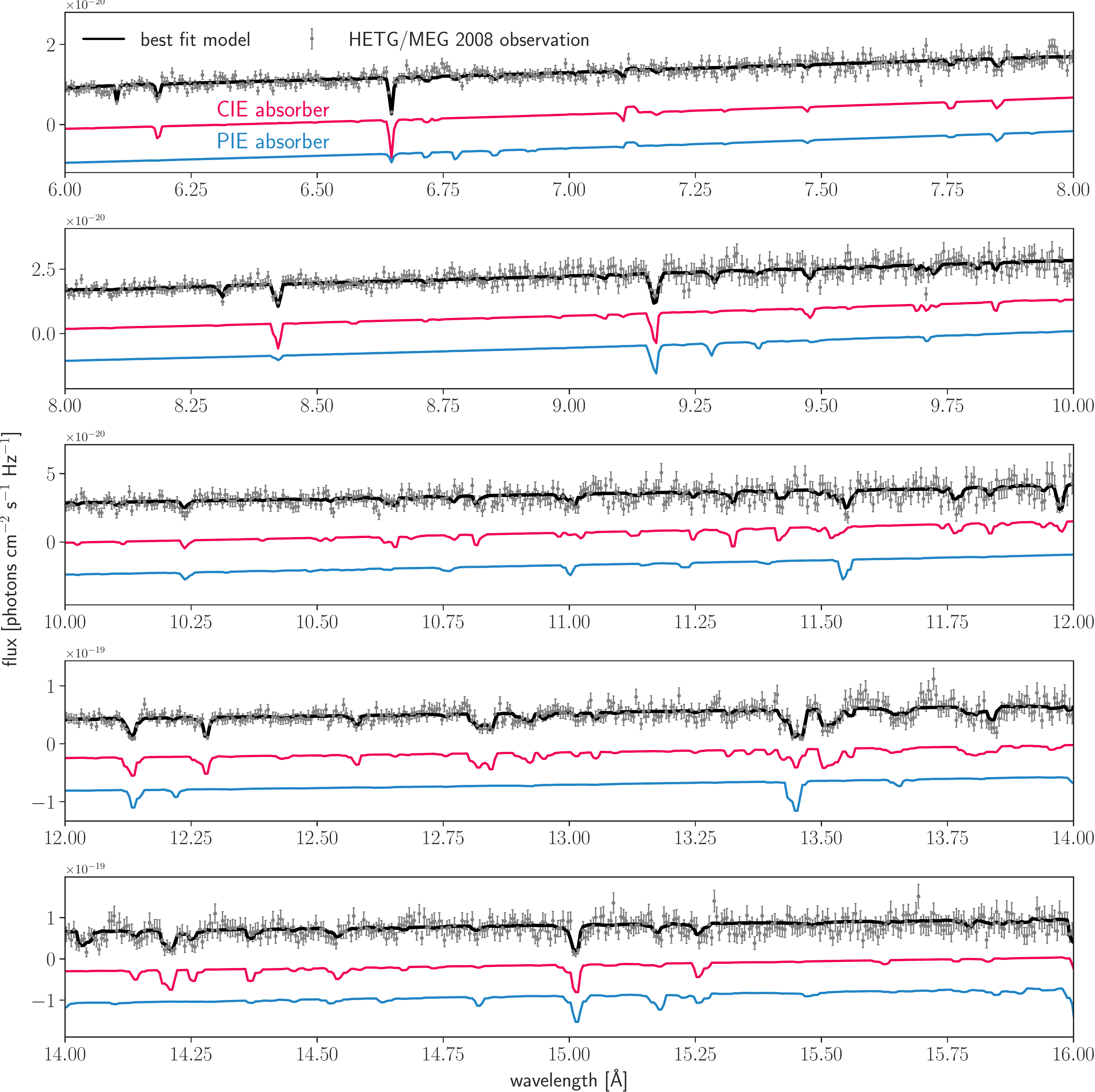}

\caption[CIE vs PIE FAST1 absorber]{The 2008 HETG/MEG spectrum is shown in grey with the most significant absorber in both data sets (FAST1) shown in red.  This component is collisionally ionized (CIE). If we did not consider collisional absorption, photoionized absorber would have been fitted, which we show in blue. We can see that the PIE absorber fits far fewer features, especially e.g.  around 12.8\AA\, 13.55\AA\ and 14.25\AA. This is expected, since the CIE model is preferred by $\Delta$DIC=189.   }
\label{fig:ciepiespec}
\end{minipage}
\end{figure*}

The most statistically significant absorber in both epochs, FAST1, is purely collisionally ionized. The evidence is overwhelming, with the CIE model being preferred to PIE model by $\Delta$DIC=189 in 2008 and $\Delta$DIC=203 in 2016 data (cf.~Table~\ref{tab:scale}). To illustrate this difference, we show best fit CIE and PIE absorbers to the spectrum in Figure~\ref{fig:ciepiespec}.  It is clear that many significant line complexes cannot be explained with a single PIE model, e.g. around 12.8\AA\, 13.55\AA\ and 14.25\AA. It appears that the CIE model produces a lot more Fe L-shell lines (in particular Fe~XVII-XVIII) under conditions that match the H-like and He-like ions of O, Ne, Mg and Si.

\begin{figure*}
\begin{minipage}{180mm}
\centering
\includegraphics[width=0.48\columnwidth,trim={0mm 0mm 0mm 0mm},clip]{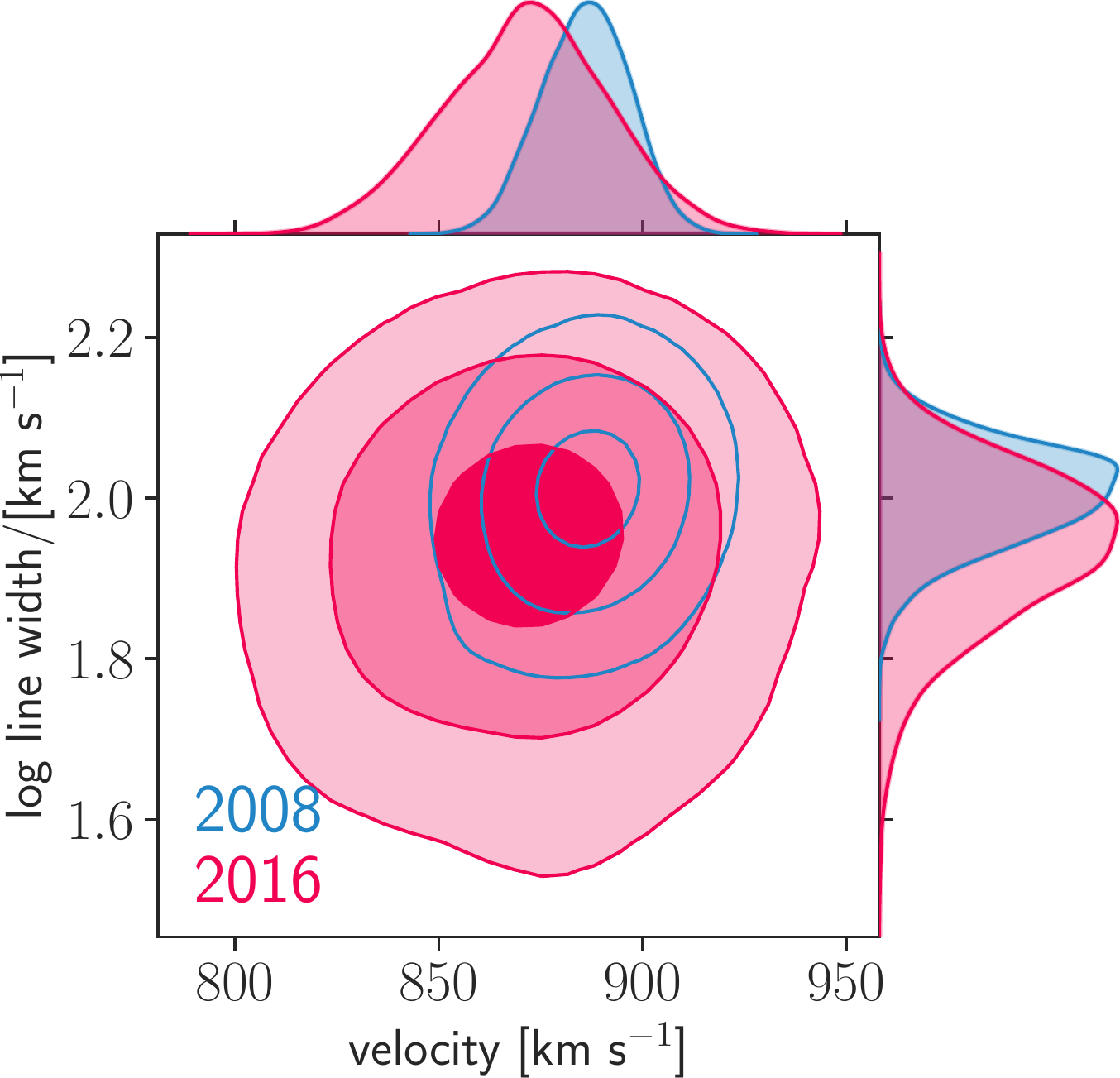}
\includegraphics[width=0.48\columnwidth,trim={0mm 0mm 0mm 0mm},clip]{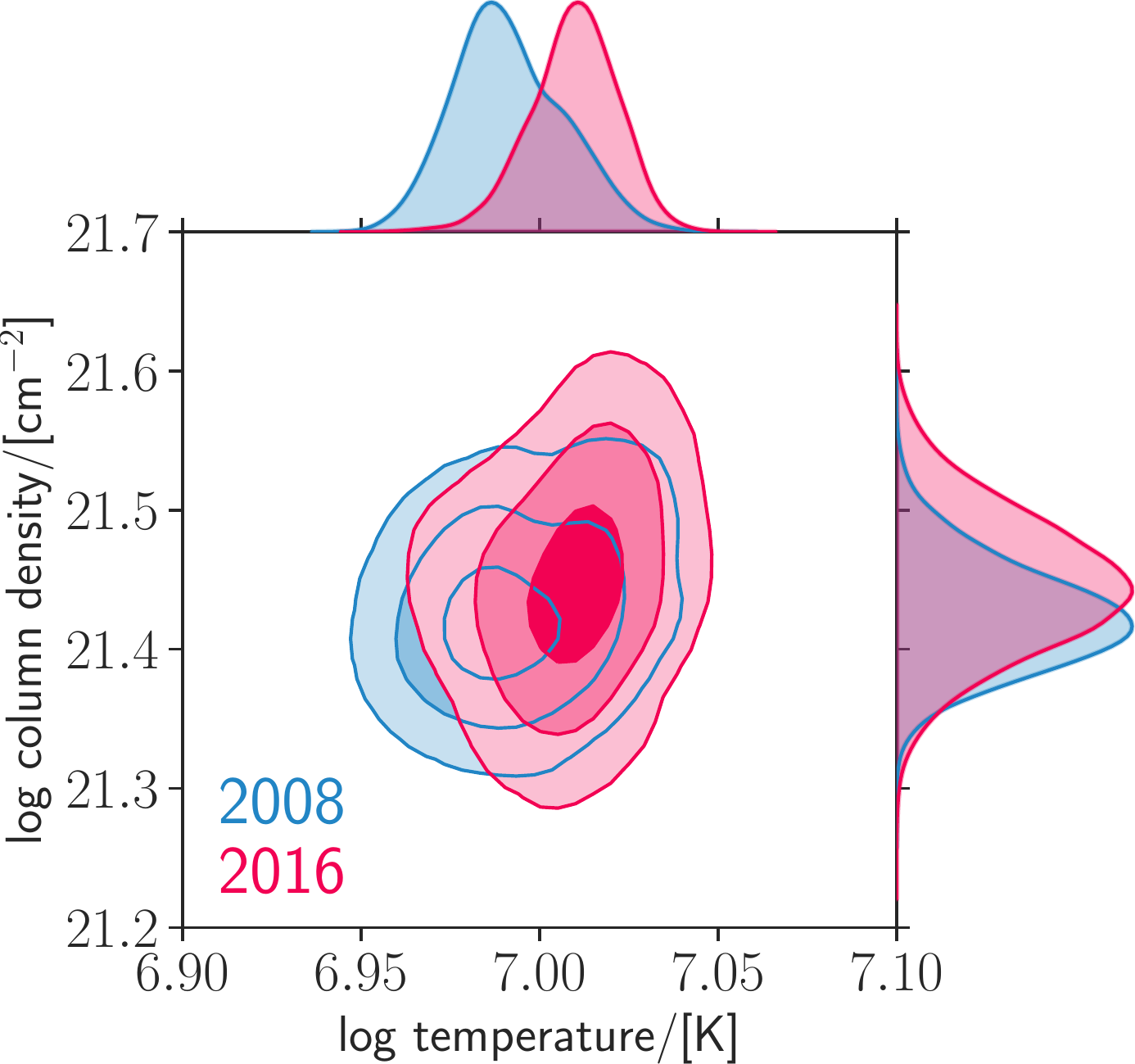}

\caption[2D best fit posteriors for all components]{2D posteriors on physical parameters of the FAST1 absorber, which is purely collisionally ionized. The 2008 epoch is shown in blue, and the 2016 is shown in red. This absorber, detected independently in both data sets, remains remarkably constant between observations. Contours are 68\%, 95\% and 99\% credible intervals.}
\label{fig:mostcie}
\end{minipage}
\end{figure*}

Furthermore, this absorber is independently detected in both epochs and remarkably consistent between them. We show a close up on its derived properties in Figure~\ref{fig:mostcie}. The temperature, velocity, line width and column density are identical in both observations.  

The origin of this absorption must be intrinsic to NGC 4051. The measured column density of this absorber, log($N_H$/[cm$^{-2}$])=21.4$^{+0.04}_{-0.03}$ and 21.4$^{+0.1}_{-0.04}$ in 2008 and 2016, is almost two orders of magnitude higher than the highest column densities measured for any ionized phases of the ISM \citep[e.g.][]{Gatuzz2018}. The derived temperature, log($T$/[K])=6.99$^{+0.02}_{-0.01}$ and 7.01$^{+0.01}_{-0.01}$ in 2008 and 2016, is an order of magnitude higher than the virial temperature of the Milky Way, as measured in X-ray absorption  \citep[e.g.][]{Gatuzz2018} and emission \citep{Nakashima2018, 2020NatAs...4.1072K}. 

The presence of this collisionally ionized outflow is not unexpected. While we detect it for the first time in a self-consistent analysis of absorption features, previous studies of NGC~4051 have also found evidence of shocked gas in this system using lower energy emission lines. \cite{Pounds2011} studied RGS spectra (which cover softer energies than HETG, $\sim$0.3-1.7~keV) in which they detected signatures of post-shock gas cooling. A similar conclusion was also drawn by \cite{Pounds2013}, who analyzed spectra from the whole X-ray band, simultaneously fitting  RGS and  EPIC (XMM-\textit{Newton} CCD) data. In addition, a number of other components in our analysis are potentially also in CIE, since the data cannopt distinguish between a PIE and CIE models (which we discuss in Section~\ref{sec:disxiden}). 

Signatures of collisionally ionized gas are also seen in a number of other local Seyfert galaxies.  CCD studies utilizing \textit{Chandra}'s high spatial resolution reveal the presence of collisionally ionized shocked gas at $\sim$100~pc scales via its X-ray emission in the cores of obscured local Seyferts (NGC~4151, \citealt{Wang2011}; Mrk~573, \citealt{Paggi2012}; Mrk~3 \citealt{Bogdan2017}; NGC~3393 \citealt{Maksym2019}). The gas temperatures measured in these works are of a similar order to our measurement. Interestingly,  collisional processes and shocks in AGN outflows have also recently been detected at higher redshifts in a large sample of quasars \citep{Mas-Ribas2019}. 

Because for gas in CIE it is not possible to estimate the wind location with our data (cf.~Section~\ref{sec:location}), we cannot determine the origin of this gas precisely. For instance, this gas could be dense enough for collisional processes to dominate over photoionization, which would suggest that it is close to the black hole \citep[cf.][]{futureHighDen}. However, since the component is stable and has relatively low velocity, this is an unlikely explanation. 

Another possibility is that this outflow is shocked when interacting with intragalacic medium, which is what could be happening in type 2 AGN discussed above. To test the reliability of this scenario, we ran a simple shock model of \cite{1979ApJS...39....1R} for an 880~km~s$^{-1}$ shock. We found that the gas heats up to $\sim$10$^7$~K and then cools down to $\sim$10$^5$~K over a column density of $\sim$10$^{21}$~cm$^{-2}$. This implies that the cooling flow behind such a shock is close to ionization equilibrium, but a range of temperatures are present, with lower temperatures having lower column densities. The observed high column density of this absorber suggest that the shock has been there for at least one cooling time. 

\subsubsection{Milky Way Absorption}
\label{sec:dismw}

The absorption by the hot halo of the Milky Way is reported on separately \citep[][]{futureMW}. However, we caution that the absorption features associated with this component, especially the 13.45 \AA\ NeIX and FeXVII blended absorption, are significant, and if not accounted for may be wrongly fitted by photoionized absorption models. 

\subsection{Outflow location}
\label{sec:location}
\begin{table} 
     \centering 
        \caption[Properties of the absorbers detected in NGC~4051]{Estimates of the distance of all PIE components detected in our data in the units of gravitational radius and pc. The $^{*}$ means that our data cannot distinguish between the PIE and CIE hypotheses. \label{tab:dist}}
 \begin{tabular}{@{}ccc}  
  \hline
component & log distance/[gravitational radius]  & log distance/[pc]  \\ 
 \hline 
    \multicolumn{3}{c}{\textbf{2008 data}} \\ 
\hline 
SLOW$^{*}$ & 10$^{+1}_{-3}$ & 3$^{+1}_{-3}$\\
FAST2 & 6$^{+2}_{-1}$ &  -1$^{+2}_{-1}$ \\ 
FAST3$^{*}$ &  8$^{+1}_{-2}$ & 1$^{+1}_{-2}$\\
VFAST1 & 6.2$^{+0.8}_{-0.8}$ & -0.9$^{+0.8}_{-0.8}$\\
\hline
  \multicolumn{3}{c}{\textbf{2016 data}} \\ 
  \hline 
SLOW & 6$^{+2}_{-2}$ & -1$^{+2}_{-2}$\\
FAST2 & 6$^{+2}_{-2}$ & -1$^{+2}_{-2}$\\
FAST3 &4$^{+1}_{-1}$ & -4$^{+1}_{-1}$\\
VFAST1$^{*}$ &2.37$^{+0.01}_{-0.01}$ & -4.720$^{+0.008}_{-0.008}$\\
VFAST2$^{*}$ & 6$^{+2}_{-2}$&-1$^{+1}_{-1}$ \\
\hline
 \end{tabular}
 \end{table}

\begin{figure}
\centering
\includegraphics[width=\columnwidth,trim={0mm 0mm 0mm 0mm},clip]{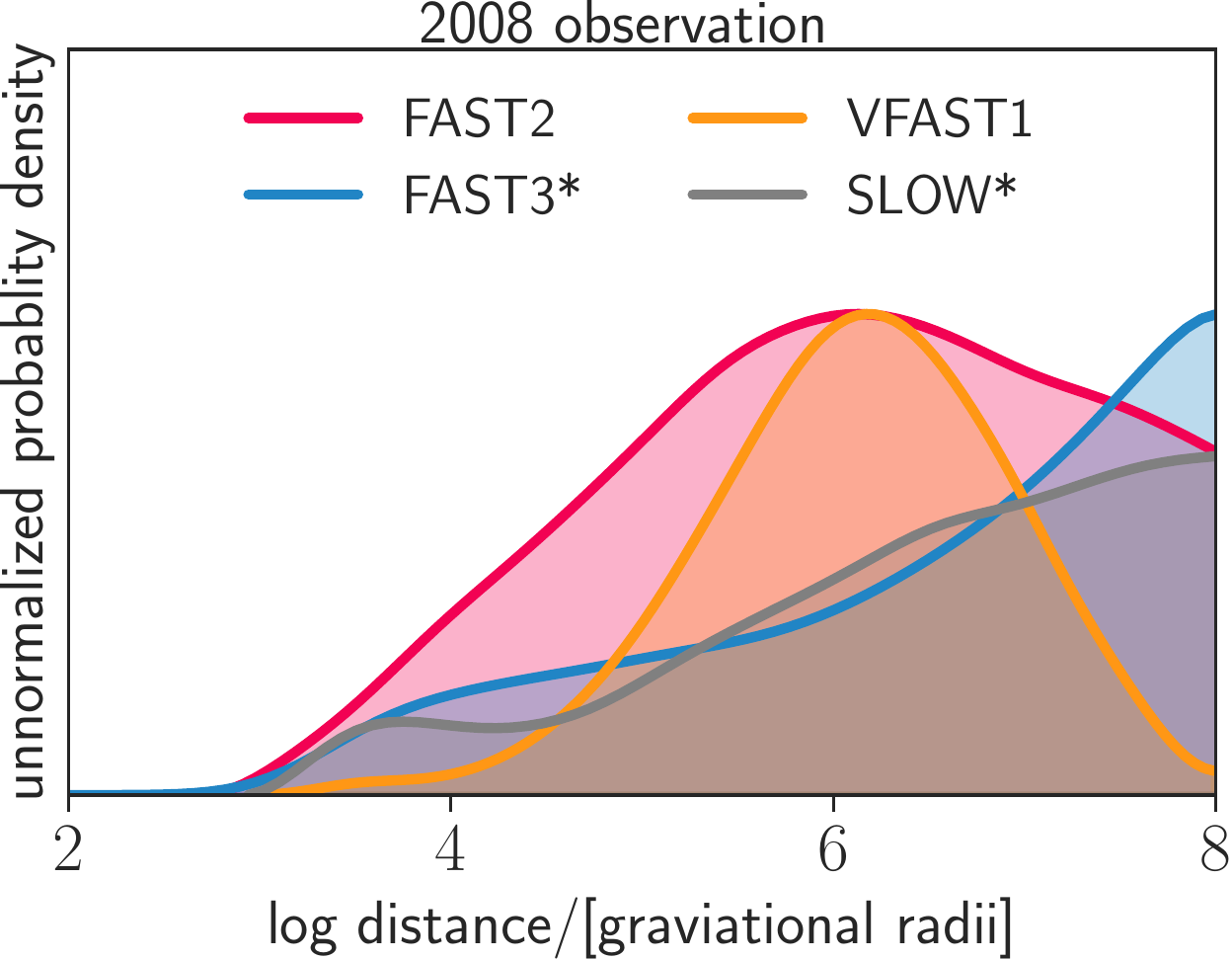}
\vspace{3ex}
\includegraphics[width=\columnwidth,trim={0mm 0mm 0mm 0mm},clip]{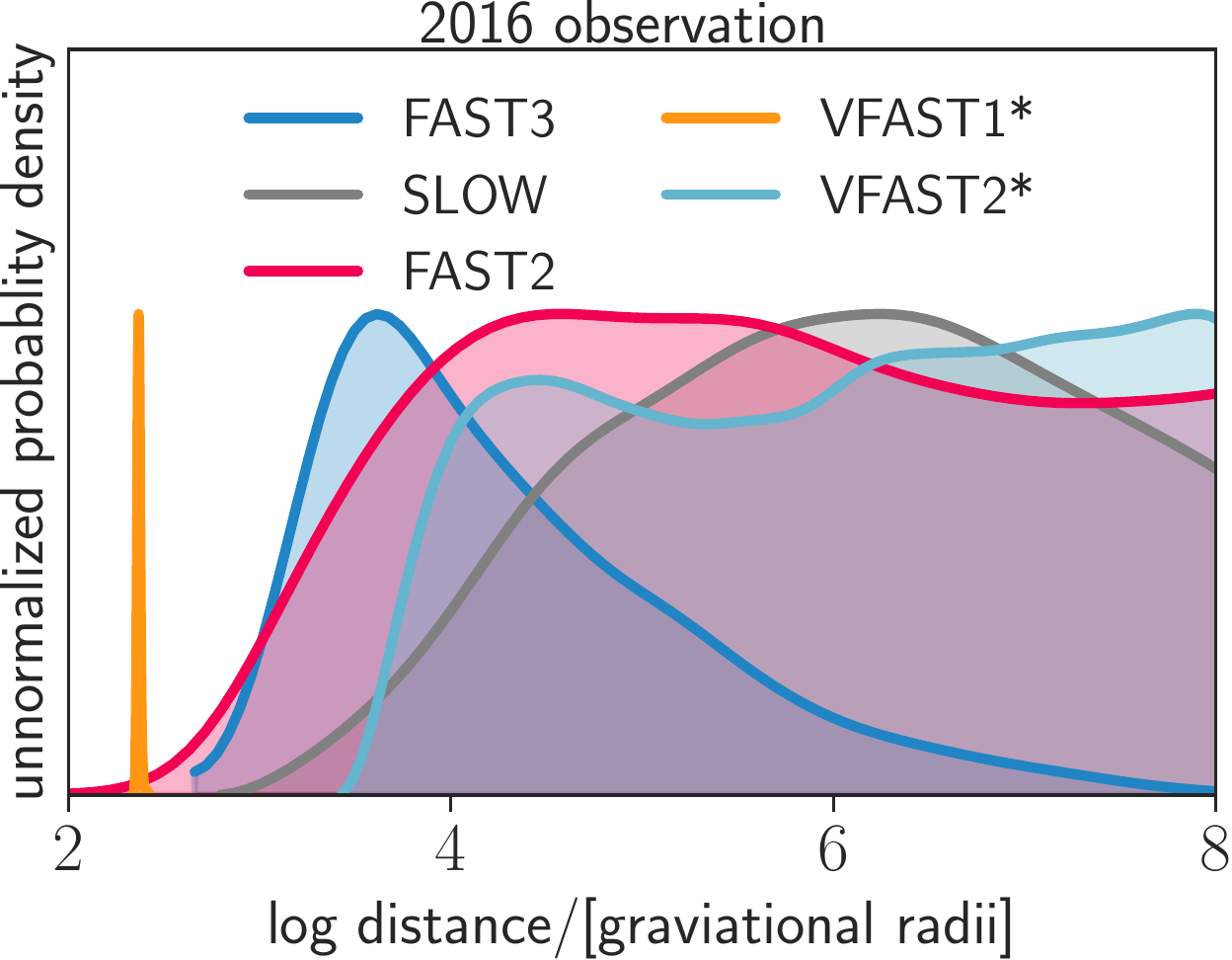}

\caption[2D best fit posteriors for all components]{Posterior distributions on the location of the photoionized outflow components in the 2008 (top) and 2016 data sets (bottom). Distances are measured in gravitational radii. The $^{*}$ notes components for with a CIE and PIE models are equally likely, making this estimate less reliable.}
\label{fig:dist}
\end{figure}

For the photoionized components we can use Equation~\ref{eq:xi} to estimate the distance to the observed clouds. Using MCMC samples and propagating errors we calculated the probability distributions on the distance to the wind components, which are shown in Figure~\ref{fig:dist} and summarized in Table~\ref{tab:dist}. With $^{*}$ we note components for which a CIE model provided an equally good fit (cf.~Section~\ref{sec:disxiden}) and the data cannot distinguish between a CIE and PIE models. In this case, our estimates may be less reliable.

Most of distance estimates have uncertainties spanning orders of magnitude, which is a consequence of large uncertainties on density measurements. The narrowest measurement is for the VFAST1 component in 2016 epoch, placing the outflow at 234 gravitational radii away from the black hole (a distance marginally consistent with the 2008 estimate of this component, which spans over 7 orders of magnitude at 3$\sigma$ level). Most components are  likely located further than 10$^4$ gravitational radii, placing them outside of the torus. 

The escape velocity of the component with the tightest constraint on distance, VFAST1/2016, is 24,000~km~s$^{-1}$. This is over a factor 2 more than its measured velocity (9700$^{+400}_{-600}$ and 9600$^{+500}_{-800}$~km~s$^{-1}$ for PIE and CIE best fits), which may mean that this outflow component would not escape the gravitational potential of the black hole. However, it is more likely that if statistical uncertainties coming from the atomic data, assumed SED and also black hole mass measurements are taken into account, these two values would be consistent. 

\subsection{Outflow power}
\label{sec:power}

\begin{figure*}
\begin{minipage}{180mm}
\centering
\includegraphics[width=0.48\columnwidth,trim={0mm 0mm 0mm 0mm},clip]{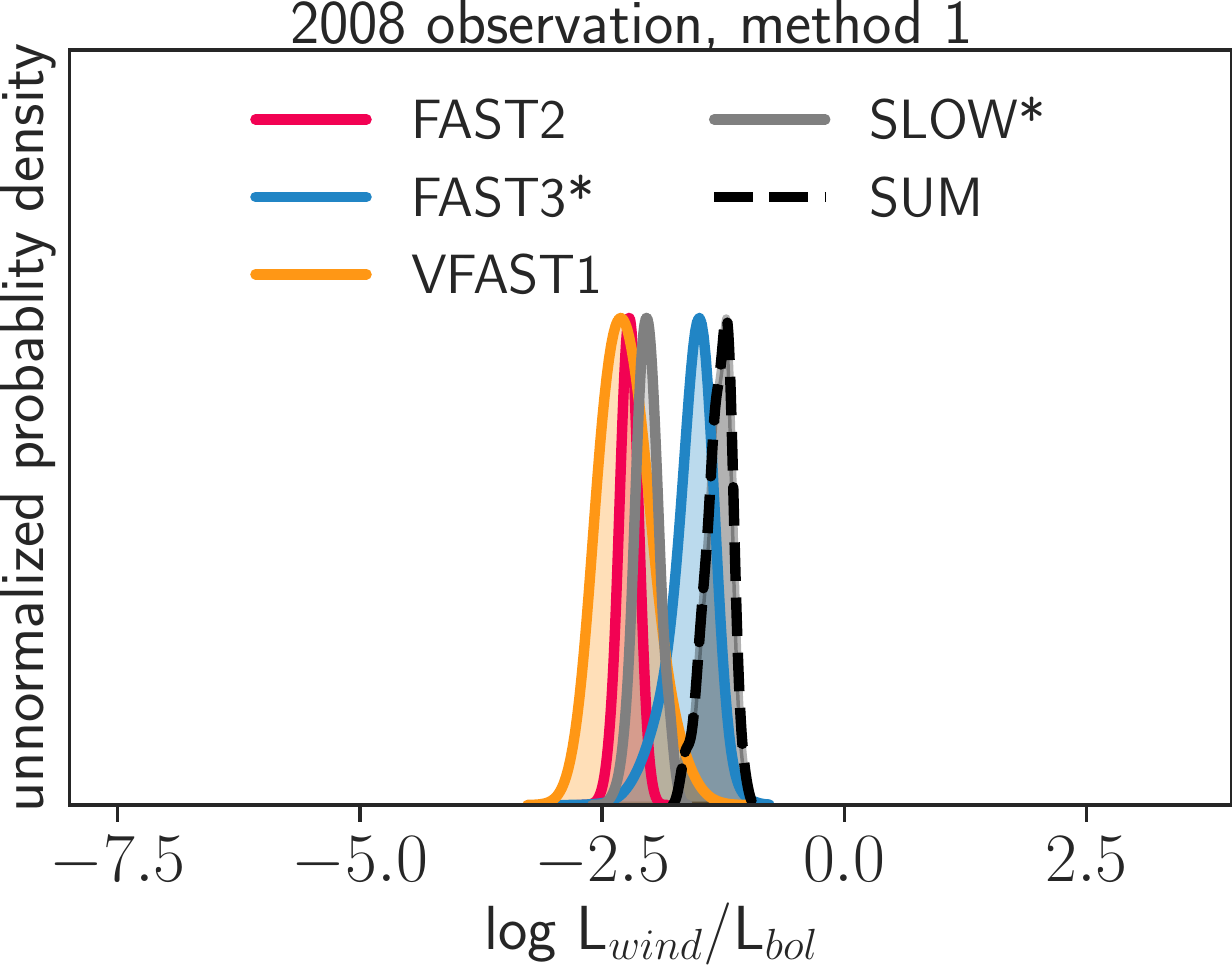}
\includegraphics[width=0.48\columnwidth,trim={0mm 0mm 0mm 0mm},clip]{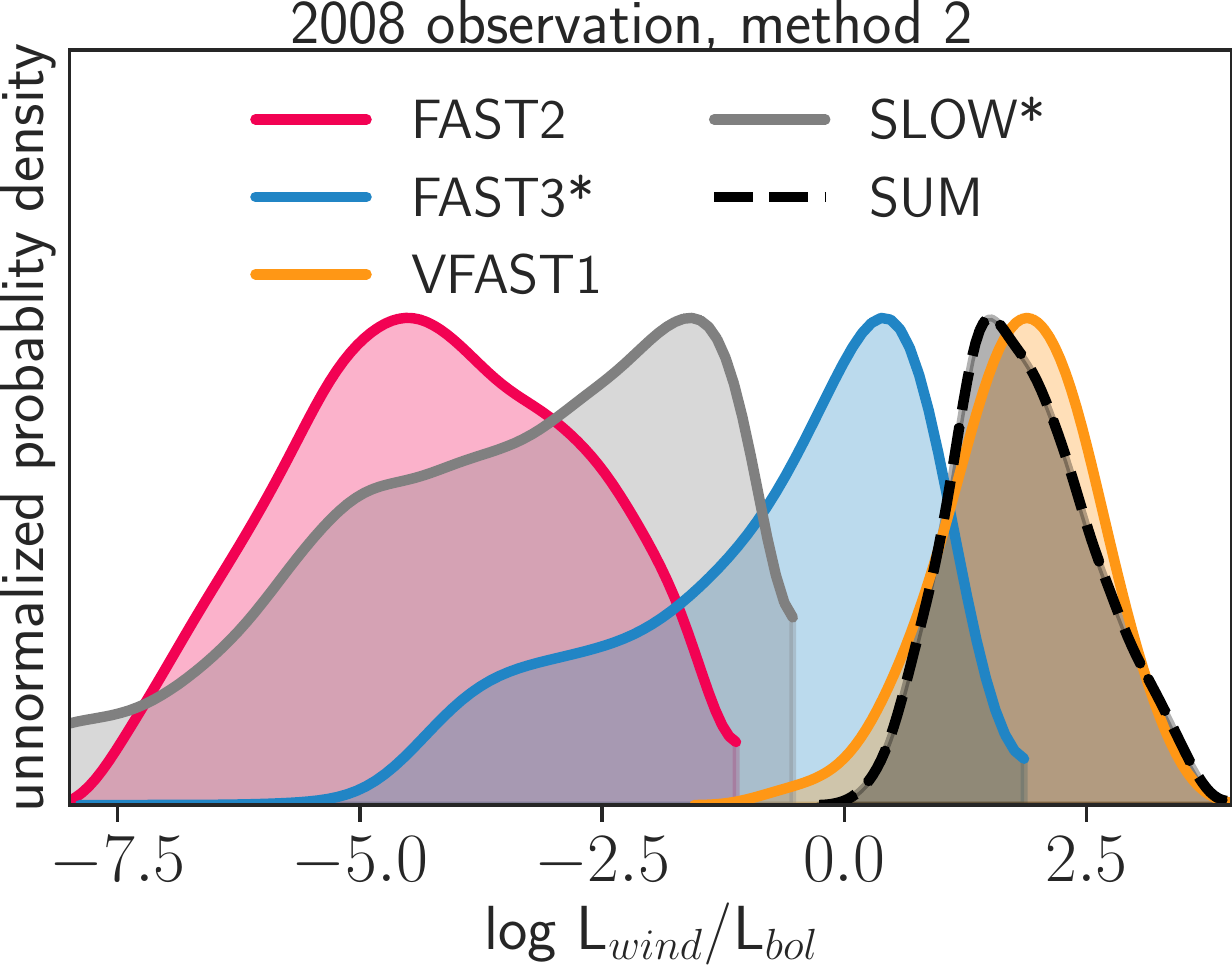}
\vspace{3ex}
\\ 
\includegraphics[width=0.48\columnwidth,trim={0mm 0mm 0mm 0mm},clip]{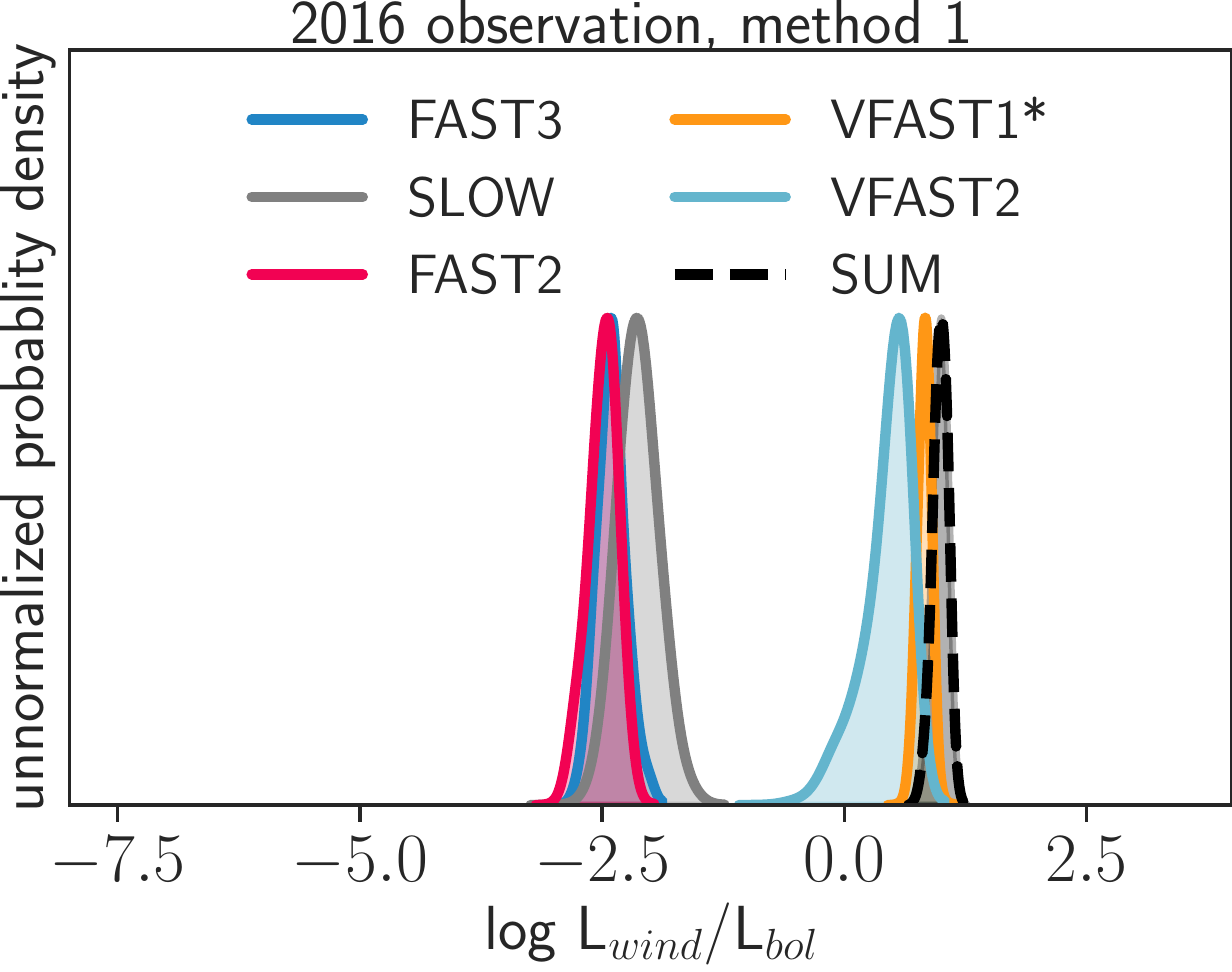}
\includegraphics[width=0.48\columnwidth,trim={0mm 0mm 0mm 0mm},clip]{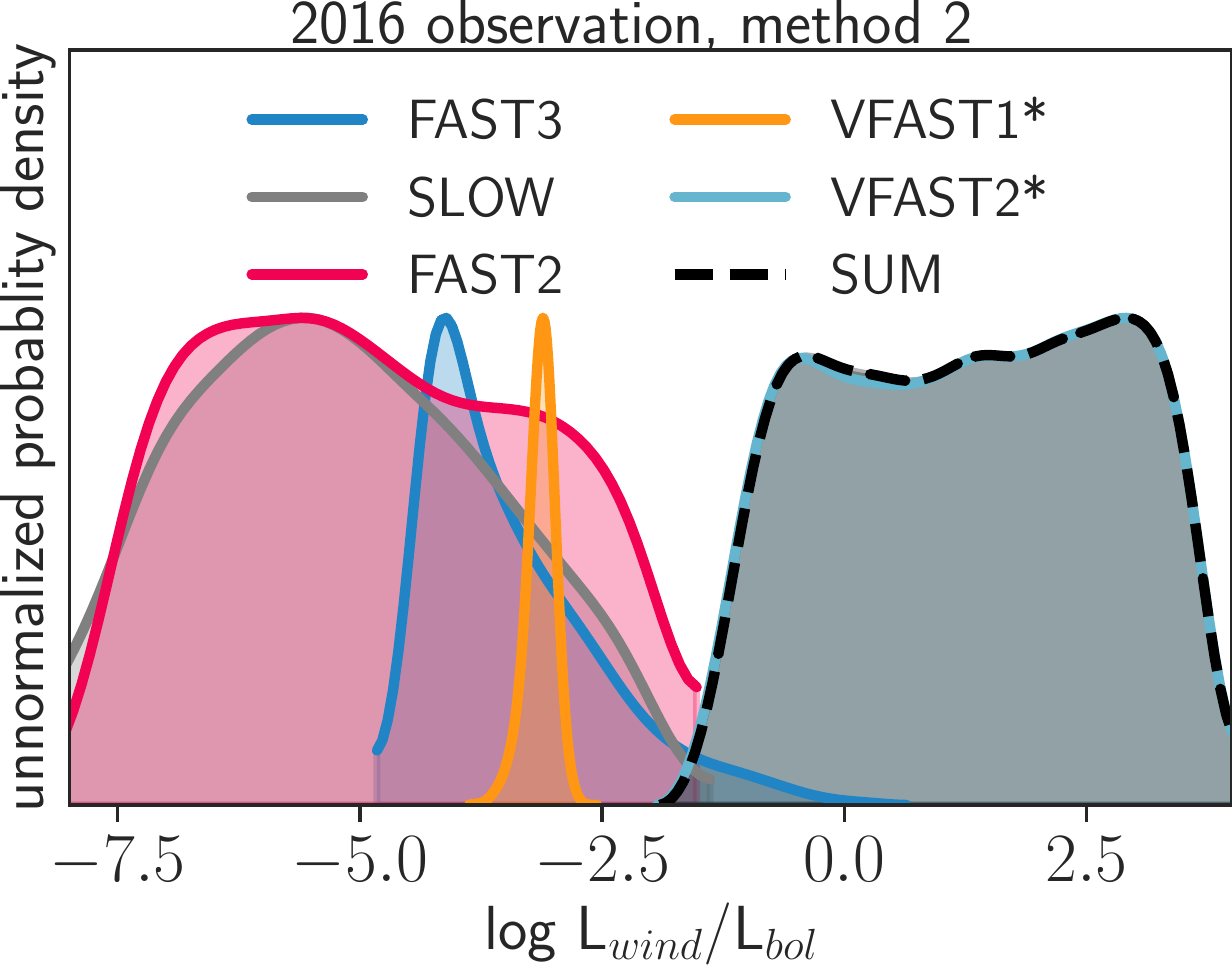}

\caption[2D best fit posteriors for all components]{Estimates of the power in each wind component in, as well as the sum over all kinetic components (dashed black line), for the 2008 and 2016 epoch (top and bottom row). We perform these calculations using two methods (left and right panels; cf.~Section~\ref{sec:power}). The $^{*}$ notes that a CIE and PIE models are equally likely, making this estimate less reliable.}
\label{fig:pow}
\end{minipage}
\end{figure*}

Using the derived physical properties of the photoionized absorbers, we can estimate their total kinetic power, $0.5 \dot{M}  v^2$, where $v$ is the outflow velocity and $\dot{M}$ is its mass loss rate. We can express mass loss rate as $\dot{M}=4  \pi \Omega \varrho r^2 v f = 4 \pi  \Omega 1.23 m_p n f $ (where $\Omega$ is the covering factor, $\varrho$ is the mass density, $n$ is the particle density, $m_p$ is proton mass, $r$ is outflow radius, and $f$ is the filling factor). To calculate outflow power, we combine this expression with Equation~\ref{eq:xi} and call it method1. Assuming spherical outflow, mass loss rate can be also expressed as $\dot{M}=4 \pi  \Omega 1.23 m_p N_H r^2 \frac{v}{r} f $ \citep[e.g.][]{2005ApJ...633..693K}, which we call method2. 

We know that $N_H=nrf$. Combined with Equation~\ref{eq:xi}, we can see that $n=\frac{N_H^2 \xi}{L_{ion} f^2}$. In pricinple, we can thus compare measured density with this expression, and estimate filling factor. Unfortunately, density measurements have large uncertainties. For majority of the absorbers, lower filling factors seem to be statistically preferred, but $f=1$ cannot be ruled out. Thus, we will assume $f=1$ for each component. This means that our power estimates should be treated as upper limits. 

We assume a covering factor of 0.5 based on population studies of outflows in Seyferts  \citep{Crenshaw2003}. For UFOs, this could be an overestimate \citep[][]{Tombesi2012}, but most of detect wind components are not relativistic. We adopt the bolometric luminosity of 10$^{43.4}$~erg~s$^{-1}$  \citep[][]{2005A&A...431..111B}, and ionizing luminosity of 10$^{42}$~erg~s$^{-1}$ \citep[][]{2011ApJ...735..107M}. Resulting outflow powers are shown for each component and as a sum over all components in Figure~\ref{fig:pow}, and summarized in Table~\ref{tab:pow}. 

We note that co-adding all components means that we assume that we are looking at different outflows rather than the same wind in different locations. In case of NGC~4051, this distinction is not significant, since the summed power is clearly dominated by a single, most powerful component. However, here we also implicitly assumed filling factor of 1. Thus, any estimates should be treated as upper limits. 

The total outflow power calculated with method 1 changes from 0.05$^{+0.02}_{-0.01}L_{bol}$ to  47$^{+454}_{-1}L_{bol}$ from 2008 to 2016, while method 2 suggest less dramatic change from $^{+2}_{-2}L_{bol}$ to 62$^{+679}_{-1}L_{bol}$. Both methods show that the outflow is becoming more powerful in the later 2016 epoch, while individual components may loose power over the same time. 

Both methods show that in both epochs the wind kinetic power is greater than 5\% of the total bolometric luminosity, and possibly even surpasses it. This would mean that outflow in NGC~4051 is important for the evolution of the host galaxy \citep[cf.][]{DiMatteo2005, Hopkins2005}.

\begin{table} 
     \centering 
        \caption[Properties of the absorbers detected in NGC~4051]{Estimates of the total outflow power summed over all kinetic components, expressed as a fraction of the NGC~4051 bolometeric luminosity, $L_{bol}$. Calculations were performed using two methods (cf.~Section~\ref{sec:power}). \label{tab:pow}}
 \begin{tabular}{@{}ccc}  
  \hline
dataset & \multicolumn{2}{c}{log (total outflow power/[$L_{bol}$])} \\
& method 1 & method 2 \\
\hline 
2008 & -1.2$^{+0.1}_{-0.2}$ & 2$^{+1}_{-1}$ \\
2016 & 1.00$^{+0.08}_{-0.08}$ & 1$^{+1}_{-1}$
\\
\hline
 \end{tabular}
 \end{table}

\subsection{Temporal variability of the outflow}
\label{sec:distemporal}

In order to show how the outflow properties change over time, we show 2D posterior distributions on column density, line width and velocity of all absorbers from both epochs in Figure~\ref{fig:2Dall}. For photoionized absorbers we show ionization and velocity in top of Figure~\ref{fig:2DPIE} and ionization and density in the bottom. For collisionally absorbers we show temperature and velocity in Figure~\ref{fig:2DCIE}. 

The three components with velocities of less than 1000~km~s$^{-1}$ remain unchanged between the two epochs. Here we are most likely observing the same physical absorber. For the remaining two components that have counterparts in both observations, FAST3 and VFAST1, some parameters differ. This could mean that either the same has evolved and changed its physical properties, or that we are seeing different, physically disconnected absorbers that move in and out of our line of sight. Longer baseline observations and comparison with theoretical models are needed to distinguish between the two scenarios. 

FAST3 absorber has significantly different velocities between the two epochs (5000$^{+300}_{-100}$~km~s$^{-1}$ for CIE and 5000$^{+300}_{-200}$~km~s$^{-1}$ for PIE best fits in 2008, and 2960$^{+40}_{-60}$~km~s$^{-1}$ in 2016). Its ionization remains constant  at approximately log($\xi/$[erg cm s$^{-1}$])$\approx$3.1 (within 1$\sigma$).  Column density is also consistent (within 1 $\sigma$, with the most probable value increasing from log $N_H$/[cm$^{-2}$]=21.5$^{+0.3}_{-0.4}$ to 22.0$^{+0.2}_{-0.2}$. Density is consistent within 2$\sigma$, but the most probable value increased from log($n/$[cm$^{-3}$])=-0.9$^{+4}_{-0.01}$ to 9$^{+1}_{-2}$. As a consequence, the location of this component in 2016 is better constrained, hinting that the component could be closer to the black hole in 2016 compared to 2008 (this is also supported by the potential decrease in line width, cf.~Section~\ref{sec:sigma}). 

Put together, this may mean that we are seeing  different parcels of gas, which are still a part of the same kinetic component of the outflow, as they move in and out of our line of sight. In 2008 we see part of the wind component that is faster, more powerful, and further from the black hole, while in 2016 we see part that is closer to the black hole and has not accelerated yet. 

\begin{figure*}
\begin{minipage}{180mm}
\centering
\includegraphics[width=\columnwidth,trim={0mm 0mm 0mm 0mm},clip]{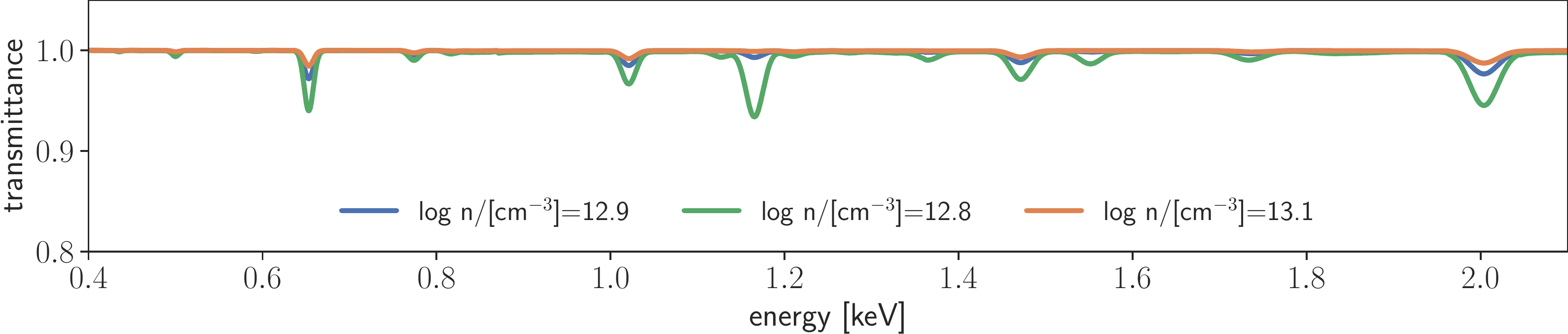}
\includegraphics[width=\columnwidth,trim={0mm 0mm 0mm 0mm},clip]{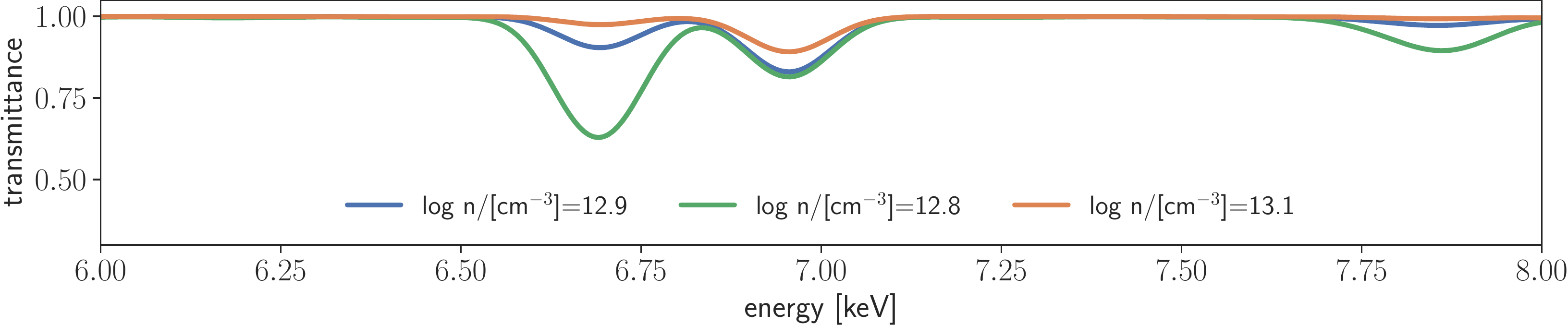}
\caption{Theoretical transmittance (ratio of the incident to transmitted radiation) of high density VFAST1/2016-like components, where density is varied around it's best fit value from our fit. Small changes in density result in significant changes of the absorption by the gas, which is why density of the VFAST1/2016 is so tightly constrained. We refer the reader to our future work on this high density regime and its ability to measure outflow density \citep[][]{futureHighDen}.} 
\label{fig:densmodel}
\end{minipage}
\end{figure*}

VFAST1 absorber remains at the same velocity in 2008 and 2016 (10000$^{+2000}_{-1000}$~km~s$^{-1}$ and 9700$^{+400}_{-600}$~km~s$^{-1}$, respectively). Ionization decreased substantially, from log($\xi/$[erg cm s$^{-1}$])=4.99$^{+0.02}_{-0.26}$ to log($\xi/$[erg cm s$^{-1}$])=1.52$^{+0.03}_{-0.02}$, and density increased significantly, from log($n/$[cm$^{-3}$])=2$^{+1}_{-2}$ to log($n/$[cm$^{-3}$])=13.0$^{+0.01}_{-0.02}$ for the PIE fit. This component is detected as purely photonized in 2008, but in 2016 the data cannot distinguish between CIE and PIE models. Its location estimate changes, placing this component orders of magnitude closer to the black hole in 2016 compared to 2008. 

In this case, it is likely we are looking  at the same wind component in two different snapshots of time. The drop in column density could suggest that some material moved out of the line of sight, enabling us more clearly to see the closer, denser, inner parts of this wind. This could also suggest that the outer layers of the wind are clumpy, or even tendril-like.

While the changes in the outflow are not dramatic, they are significant enough to support observational campaigns of this source with high throughput high spectral resolution X-ray observatories over timescales of years, in order to further understand the structure of the outflow and compare it to models of wind physics.

\subsection{Density constraints}

We have obtained density constraints for all photoionized absorbers. For all but one absorbers (VFAST1/2016) these constraints span orders of magnitude. This is expected, since moderate density does not influence photoionization equilibrium, which is a function of ionization parameter for these low to moderate densities. Note the exception to this are narrow ranges of the parameter space, where density sensitive lines are important and can be used as a diagnostic \citep[][]{2017A&A...607A.100M}, or density sensitive line ratios \citep[e.g.][]{2003ApJ...588L.101M, 2008ApJ...680.1359M}.

However, in the low and high density regime, the photoionized equilibrium starts to depend on density in addition to the ionization parameter. This why high densities can be robustly excluded for majority of the PIE components, and the particular density range that can be excluded depends on the ionization parameter (cf. Figure~\ref{fig:2DPIEdenxi}, where the upper bounds on the density become lower as ionization parameter increases, creating high density exclusion zone). This is also why for the highest density component in our data, log($n/$[cm$^{-3}$])=13.0$^{+0.01}_{-0.02}$, the model has high constraining power and the resulting measurement is so tight. We illustrate the constraining power of the model in this case in Figure~\ref{fig:densmodel}.  We discuss this dependence on density further in our future work \citep[][]{futureHighDen}.

\subsection{Line broadening} \label{sec:sigma}

For photoionized components, for which wind density is also constrained, we can compare the measured line width to the Keplerian velocity at the distance of the absorber. If the broadening comes from Keplerian rotation, it will be order of $r/R$ times the Keplerian speed, where $r$ is the size of the emitting region and $R$ is the distance of the absorber \citep{2020ApJ...899L..16T}.

We find that for most of the photoionized absorbers the Keplerian broadening and measured line width agree well (within the large uncertainties), so we cannot make any claims about the size of the emitting region. The exception is VFAST1 component, where the measured broadening is  larger than the Keplerian estimate in 2008 by almost 2 orders of magnitude, and lower by an order of magnitude in 2016 (albeit with large uncertainties). The lower measurement could suggest that a significant fraction of rotational velocity is perpendicular to our line of sight; higher could mean additional velocity shear within the component itself. However, measurements with smaller uncertainties and theoretical models are needed to use the line width as a useful physical probe. 

\subsection{Soft excess and relativistic disk reflection}\label{sec:disrel}

We find that in the 2016 spectrum, soft excess is fully accounted for by the relativistic blurred reflection in the Fe-L band, while in the 2008 spectra an additional phenomenological \textit{diskbb}  is statistically preferred. Without the additional \textit{diskbb}, the power law index gets softer  and normalization gets higher, as this component is trying to account for some of the soft emission. Including the \textit{diskbb} brings the \textit{powerlaw} down, making it fully consistent with the 2016 measurement (and with other analyses of the same dataset, e.g. \citealt{King2012}).

In all models considered, the black hole spin is constrained at its maximal value, the disk density fitted pegs at the maximum value allowed by the model, 10$^{19}$~cm$^{-3}$ (such high disk densities are plausible for relatively low mass SMBHs,  like the one in NGC~4051). It must be emphasized that the upper limit on the disk density in the \emph{relxillD} model is not physically motivated, but stems from the lack of atomic data that underlies the calculations for such high density plasma \citep{Garcia2016}. As a consequence, we cannot say anything on whether there are any changes in disk density and mass accretion rates. 

Therefore, it is likely that the soft excess could  still be fully explained by high density disk relativistic reflection, if the model used was complete.  A number of recent works have also shown that the high disk density blurred relativistic models can account for most if not all of soft excess radiation in Seyfert galaxies \citep{Mallick2018, Jiang2019, Garcia2019}.

The inclination constrained by \textit{relxillD} is consistent between the two epochs,  (and essentially identical for both data sets in fits without the phenomenological \textit{diskbb}) at 56$^{+1\: \circ}_{-4}$ and 49$^{+1\:\circ}_{-2}$ i 2008 and 2016 respectively. It is also consistent with other, independent measurements of NGC~4051 inclination, such as observations if its jet \citep[40-60$^\circ$; ][]{Maitra2011} and Narrow Line Region \citep[$50^\circ\pm11^\circ$; ][]{Christopoulou1997}. 

The disk ionization changes slightly between the two epochs, which is expected, since the ionization is variable on light crossing timescales. 

\subsection{Advantages and limitations of our approach} \label{sec:disgrid}
\label{sec:limitations}

A thorough exploration of the parameter space is always challenging, and in high dimensional parameter spaces it is typically impossible to be certain that the global maximum in the likelihood space has been found. Our approach has been to take advantage of high performance computing in order to carefully search the parameter space of AGN winds physical properties. We utilized the computing power to build an extensive, finely populated grid of photoionization models, and to also perform  self-consistent fits. 

The advantage of this approach is that we were able to probe ranges of parameter space typically excluded from analyses, such as very high densities, and orders of magnitude in line width. Additionally, not being restricted to a grid, we avoid biases from interpolation between grid points.

The downside to our approach is that it requires nontrivial computing resources ($\sim$months on a supercomputer). Currently the number of data sets requiring more sophisticated approaches is limited, since high signal to noise spectra require extremely long integration times (>0.5~Ms). However, these will soon be common with the launch of new generation of X-ray spectrographs, such as X-ray Imaging and Spectroscopy Mission (XRISM; 2023) and Advanced Telescope for High Energy Astrophysics (Athena; 2032).

\section{Conclusions}\label{sec:conclusions}

We have analyzed two deep \textit{Chandra} HETG  spectra of Seyfert galaxy NGC~4051: a 701~ks observation  from 2016 and a ~314~ks observation from 2008. We constructed a fully self-consistent Bayesian framework which allowed us for robust model comparison and utilization all of predictive power of the data. 

Using the DIC, an approximation of Bayesian evidence, we constructed continuum emission model. Interestingly, most of the 2008 and all of the 2016 soft excess emission can be accounted for by relativistic blurred reflection from a high density ($10^{19}$~cm$^{-3}$) accretion disk. A warm corona as the origin of this emission is statistically ruled out. Our measurement of the disk inclination, $\sim$50$^{\circ}$, is consistent with constraints from jet and Narrow Line Region observations.

We then looked for absorbers that are either in collisional or photoionization equilibrium, and added them to our model based on the DIC. This resulted in a detection of six absorbers intrinsic to NGC~4051, five of which are present in both epochs. We categorize one absorber as SLOW ($~\sim400$~km~s$^{-1}$), three as FAST ($~\sim1000-5000$~km~s$^{-1}$), and two as VFAST ($~\sim10,000-30,000$~km~s$^{-1}$). 

The most statistically significant wind component is purely collisionally ionized with a temperature $T=10^7$~K and velocity $v=880$~km~s$^{-1}$, which is a first detection in absorption of a such an AGN wind. Remarkably, the physical properties of this component are precisely measured and yet stable across the two epochs. This physical state of the outflow may be achieved either through high density or shocks. However, we are unable to distinguish between these two scenarios with the current data. 

Two absorbers in 2008 and three in 2016 are  purely photoionized, while two in each observation are described by CIE and PIE models equally well. This may suggest that these components are a mixture of photoionized and collisionally ionized gas, which could be caused by high density. However, it is also likely that the current data is insufficient to distinguish between these two models. 

For one of the relativistic absorbers moving at 3\%$c$, we obtain one of the tightest density measurements to date, log~$n/$[cm$^{-3}$]=13.0$^{+0.01}_{-0.02}$, which is close to the typical accretion disk densities. We further determine that this component is located at  $\sim$240 gravitational radii away from the black hole. We note, however, that atomic data, assumed SED, and black hole mass uncertainties are all possible sources of systematic error on these measurements.

The physical properties of the three slowest components are consistent between the two epochs, while the two faster ones show signs of variability. With just two observational epochs, however, we cannot determine if these temporal changes mean that we are looking at different wind components, or the same component undergoing physical changes.

We estimated the  energetics of the NGC~4051 outflow components and found that in each epoch the outflow kinetic power exceeds 5\% of the bolometric luminosity, and may be even more powerful (however, uncertainties on these estimates are large). This suggests that the X-ray detected outflow in NGC~4051 can impact the evolution of its host galaxy.

\section*{Data availability}
Analyzed \textit{Chandra} HETG data are publicly available at NASA's High Energy Astrophysics Science Archive Research Center . Probability distributions presented in this article will be shared on reasonable request to the corresponding author.

\section*{Acknowledgements}
The authors thank the anonymous referees for comments that improved the manuscript.  AO thanks Peter van Hoof for providing advice on how to modify the Cloudy code  via the Cloudy user group, Gary Ferland for his help with interpreting Cloudy results and general guidance on photoionized plasmas, Marios Chatzikos and other Cloudy developers for creating a new capability to identify absorption features, Tim Kallman and Adam Mantz for useful discussions, and Jon Miller for useful discussions and proposing for the observations. Some of the computing for this project was performed on the Sherlock cluster. We would like to thank Stanford University and the Stanford Research Computing Center for providing computational resources and support that contributed to these research results. We thank the authors and maintainers of the following \textsc{Python} packages that have been used in this analysis: \texttt{emcee} \citep{emcee}, \texttt{schwimmbad} \citep{schwimmbad}, \texttt{corner} \citep{corner}, \texttt{matplotlib} \citep{matplotlib}, \texttt{NumPy} \citep{numpy}, and \texttt{SciPy} \citep{scipy}. This research has made use of software provided by the Chandra X-ray Center (CXC) in the application package CIAO. This research has made use of data and/or software provided by the High Energy Astrophysics Science Archive Research Center (HEASARC), which is a service of the Astrophysics Science Division at NASA/GSFC and the High Energy Astrophysics Division of the Smithsonian Astrophysical Observatory. This research has made use of the NASA/IPAC Extragalactic Database (NED), which is operated by the Jet Propulsion Laboratory, California Institute of Technology, under contract with the National Aeronautics and Space Administration. The material is based upon work supported by NASA under award number 80GSFC21M0002.

\bibliographystyle{mnras}
\bibliography{bib} 

\appendix

\section{Absorption features of individual components} \label{sec:absfigs}

\begin{figure*}
\begin{minipage}{180mm}
\centering
\includegraphics[width=\columnwidth,trim={0mm 0mm 0mm 0mm},clip]{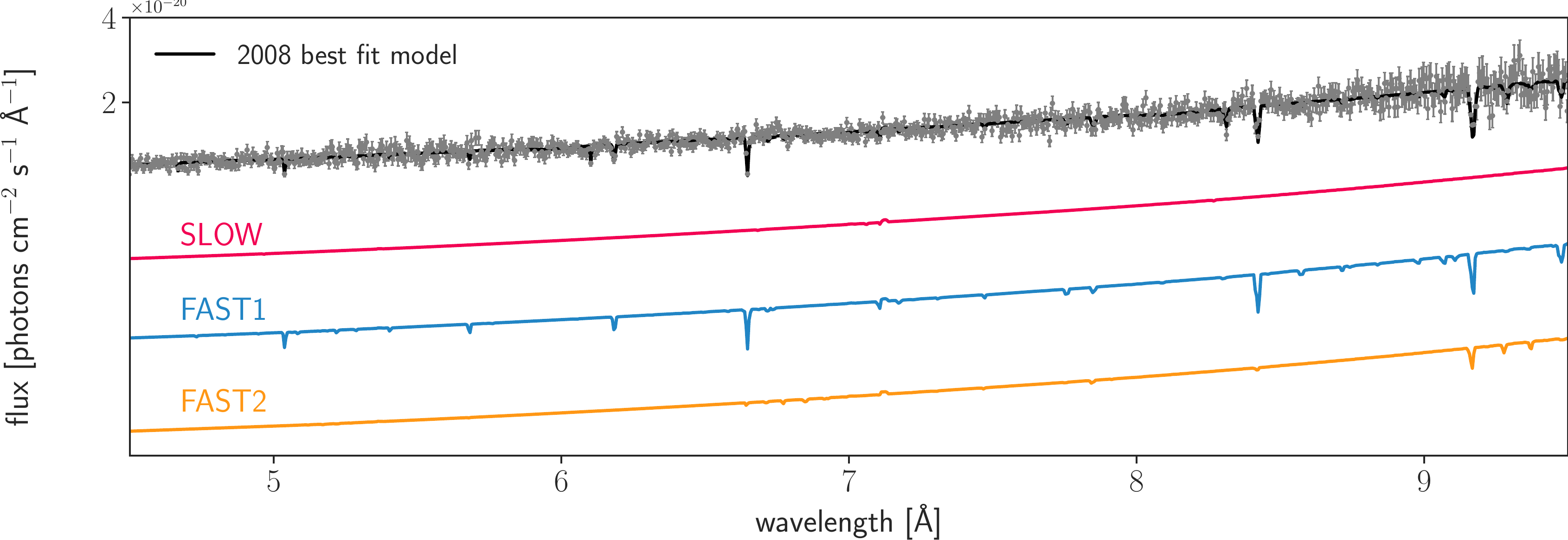}
\includegraphics[width=\columnwidth,trim={0mm 0mm 0mm 0mm},clip]{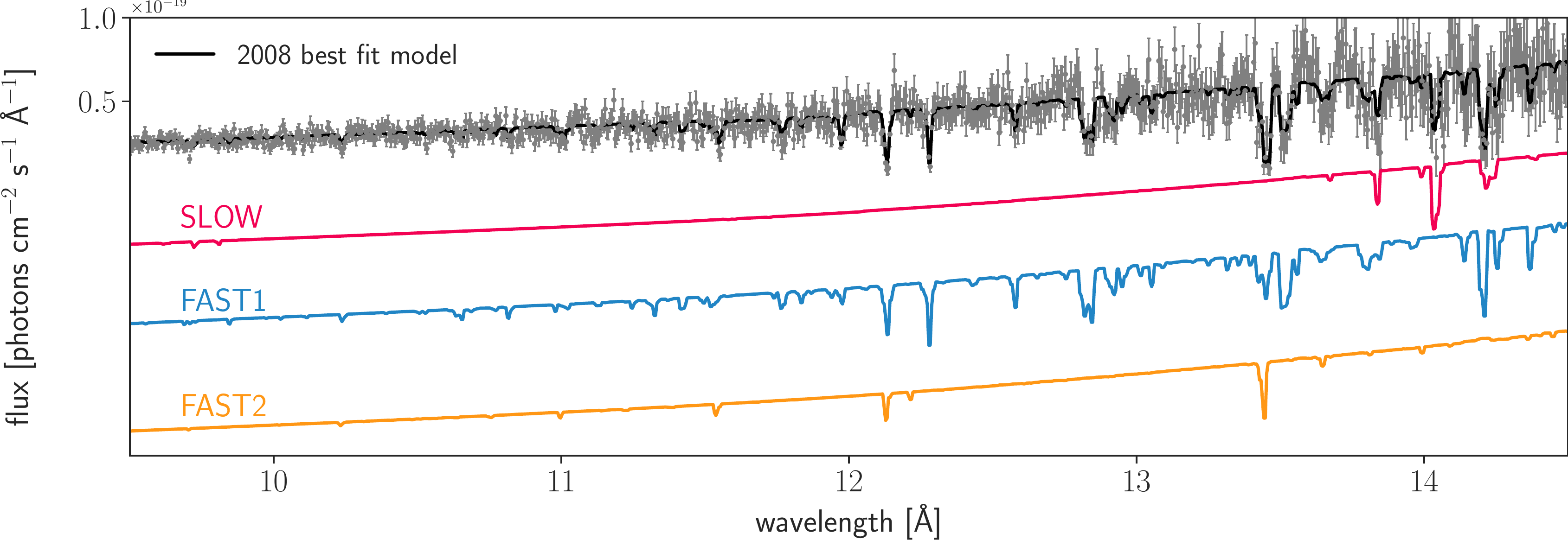}
\includegraphics[width=\columnwidth,trim={0mm 0mm 0mm 0mm},clip]{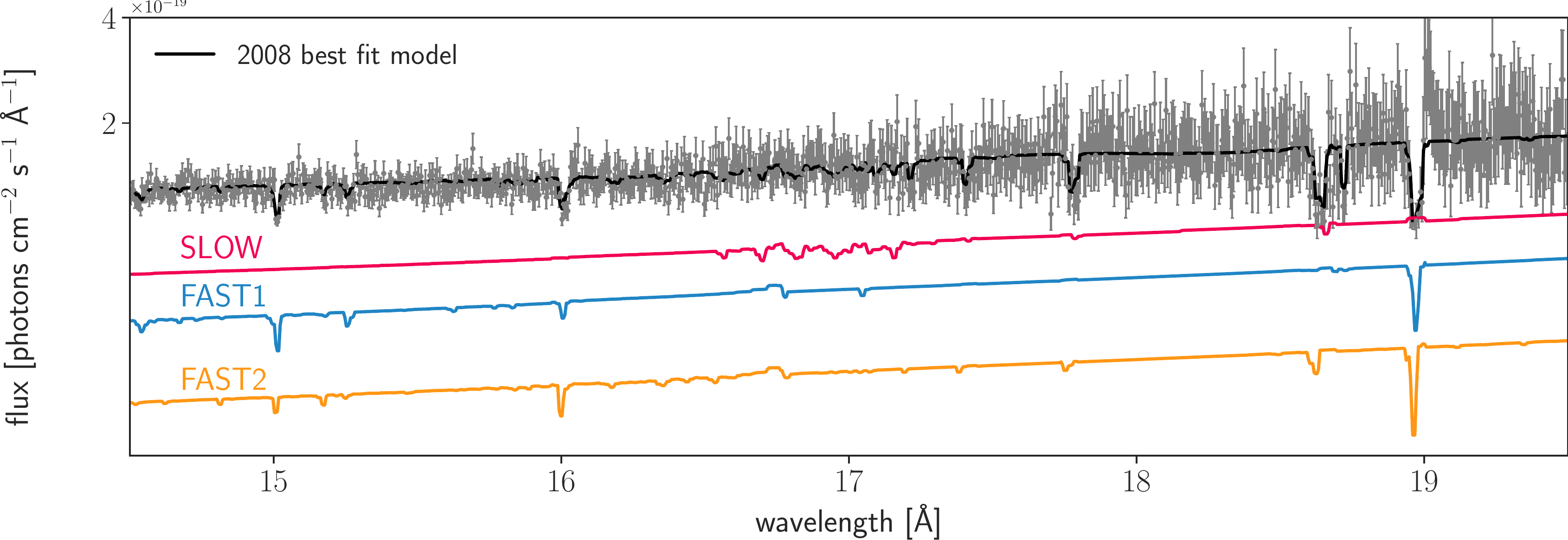}
\caption{Spectral features imprinted by narrow line width absorbers found in the 2008 data. The black line is the overall best fit model (\textit{neutral MW * hot MW * all AGN absorbers * AGN continuum}), plotted on top of the 2008 HETG/MEG data (grey). Below, narrow line width absorbers are shown individually, i.e.  \textit{neutral MW * individual absorber * AGN continuum}. For identification of the most significant individual features, please refer to Table~\ref{tab:lines08}. Several unlabelled features are from the hot halo of the Milky Way \citep[see][]{futureMW}. Figure  continues in~\ref{fig:narrowcomps08cont}.} 
\label{fig:narrowcomps08}
\end{minipage}
\end{figure*}

\begin{figure*}
\begin{minipage}{180mm}
\centering
\includegraphics[width=\columnwidth,trim={0mm 0mm 0mm 0mm},clip]{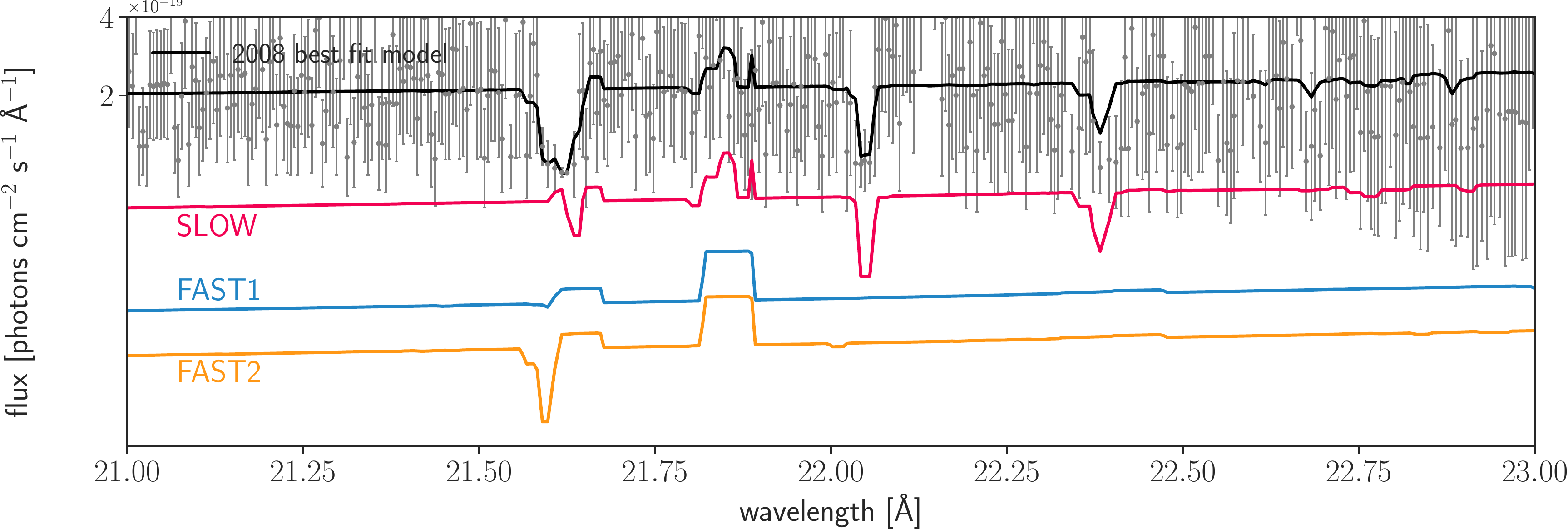}
\caption{Figure~\ref{fig:narrowcomps08} continued.}
\label{fig:narrowcomps08cont}
\end{minipage}
\end{figure*}

\begin{figure*}
\begin{minipage}{180mm}
\centering
\includegraphics[width=\columnwidth,trim={0mm 0mm 0mm 0mm},clip]{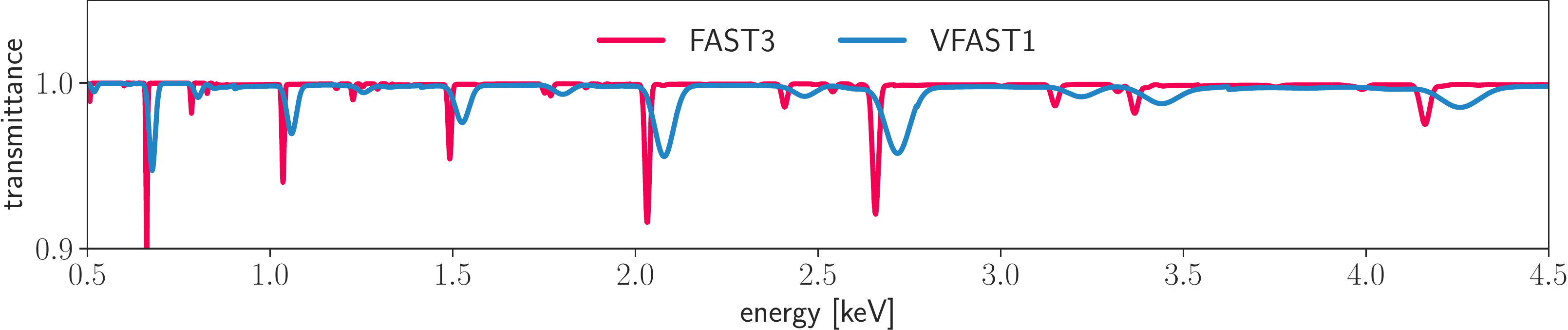}
\includegraphics[width=\columnwidth,trim={0mm 0mm 0mm 0mm},clip]{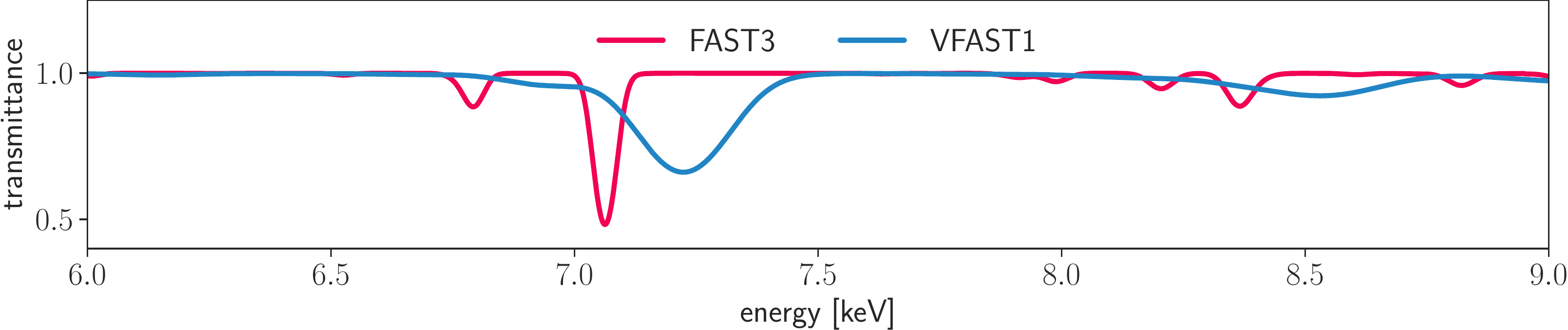}
\caption{Transmittance (ratio of the incident to transmitted radiation) for the broad line width best fit absorbers in the 2008 epoch. Due to their width, these features are not identifiable 'by eye' in high spectral resolution data. For identification of the most significant individual features, please refer to Table~\ref{tab:lines08}. } 
\label{fig:broadcomps08}
\end{minipage}
\end{figure*}

\begin{figure*}
\begin{minipage}{180mm}
\centering
\includegraphics[width=\columnwidth,trim={0mm 0mm 0mm 0mm},clip]{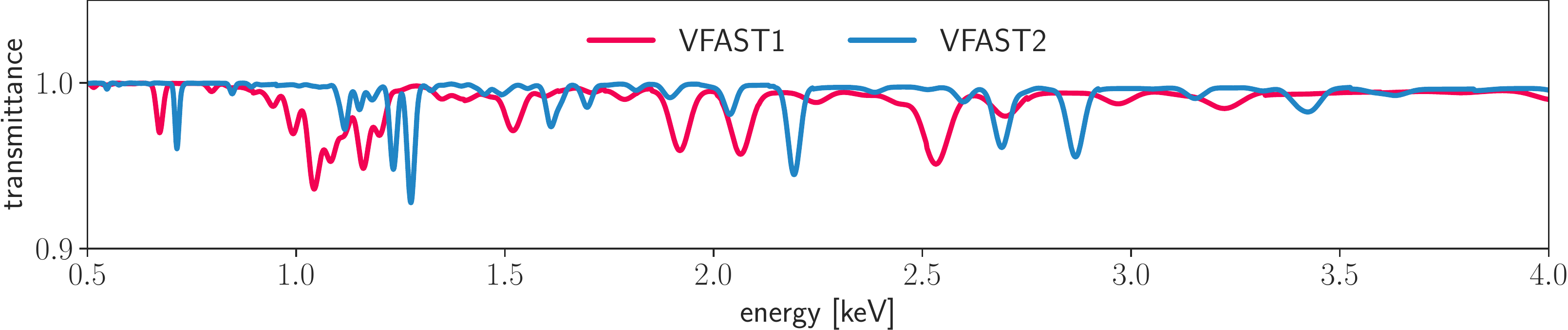}
\includegraphics[width=\columnwidth,trim={0mm 0mm 0mm 0mm},clip]{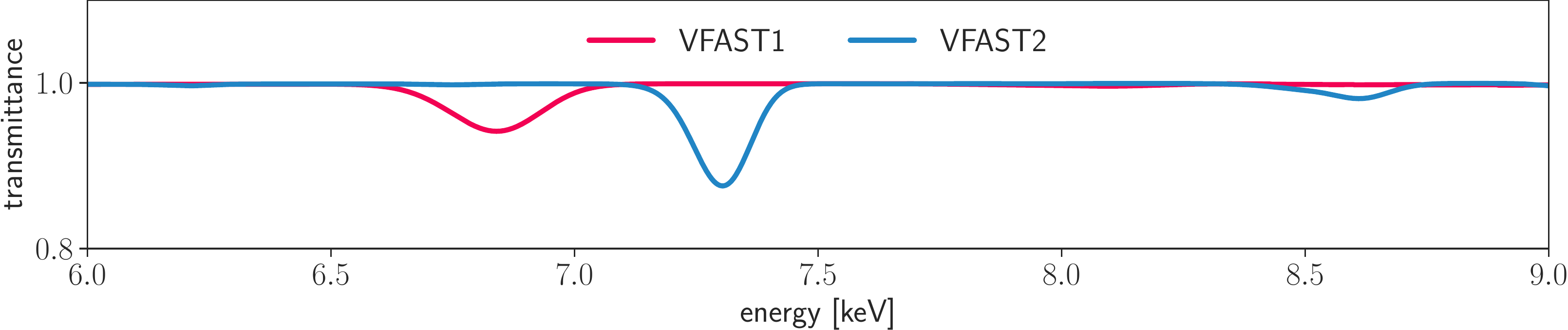}
\caption{Transmittance (ratio of the incident to transmitted radiation) for the broad line width best fit absorbers in the 2016 epoch. Due to their width, these features are not identifiable 'by eye' in high spectral resolution data. For identification of the most significant individual features, please refer to Table~\ref{tab:lines16}. } 
\label{fig:broadcomps16}
\end{minipage}
\end{figure*}

Each detected wind component imprint many features onto the spectrum. We show individual narrow line absorbers directly in the data in Figures~\ref{fig:narrowcomps08} and~\ref{fig:narrowcomps08cont} for the 2008 epoch, and in Figure~\ref{fig:broadcomps08} for the 2016 epoch. We show the best fitting models for the broad line widths absorbers with their transmittance in Figures~\ref{fig:broadcomps08} and~\ref{fig:broadcomps16} for the 2008 and 2016 datasets, respectively.

We then list most significant absorption features in each of the individual wind components in Tables~\ref{tab:lines08} and~\ref{tab:lines16} for the 2008 and 2016 epochs. These features have the highest equivalent widths and are in regions where HETG is sensitive.

\begin{figure*}
\begin{minipage}{180mm}
\centering
\includegraphics[width=\columnwidth,trim={0mm 0mm 0mm 0mm},clip]{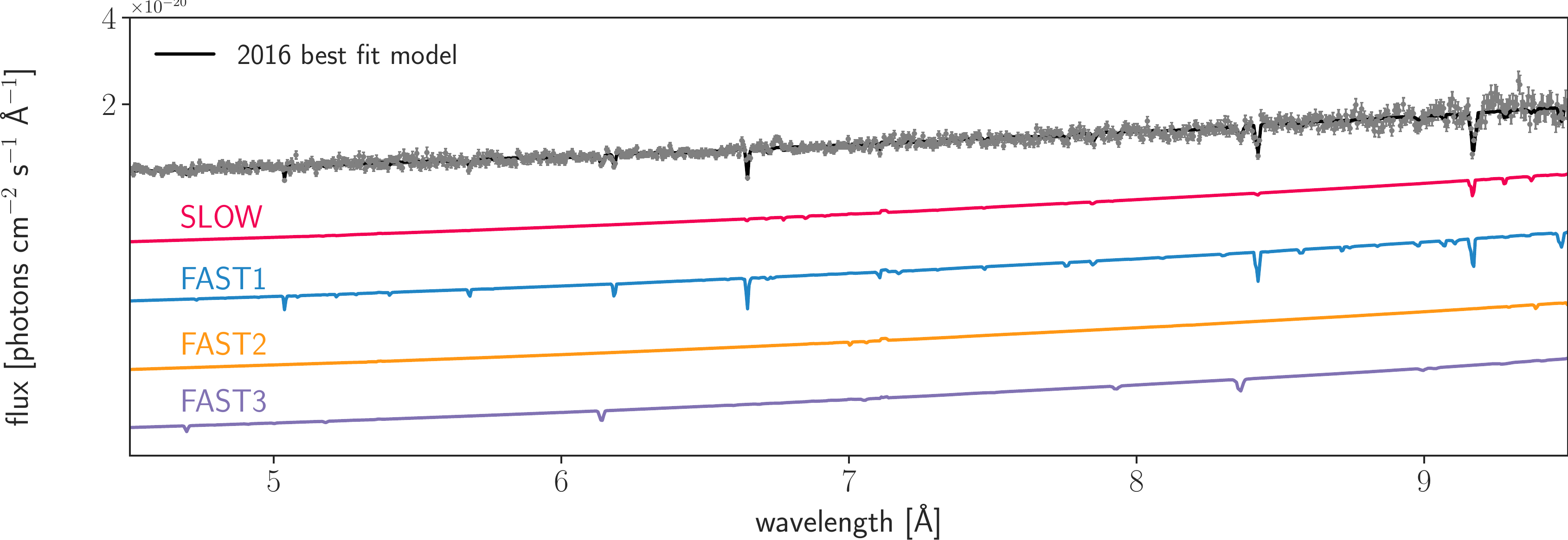}
\includegraphics[width=\columnwidth,trim={0mm 0mm 0mm 0mm},clip]{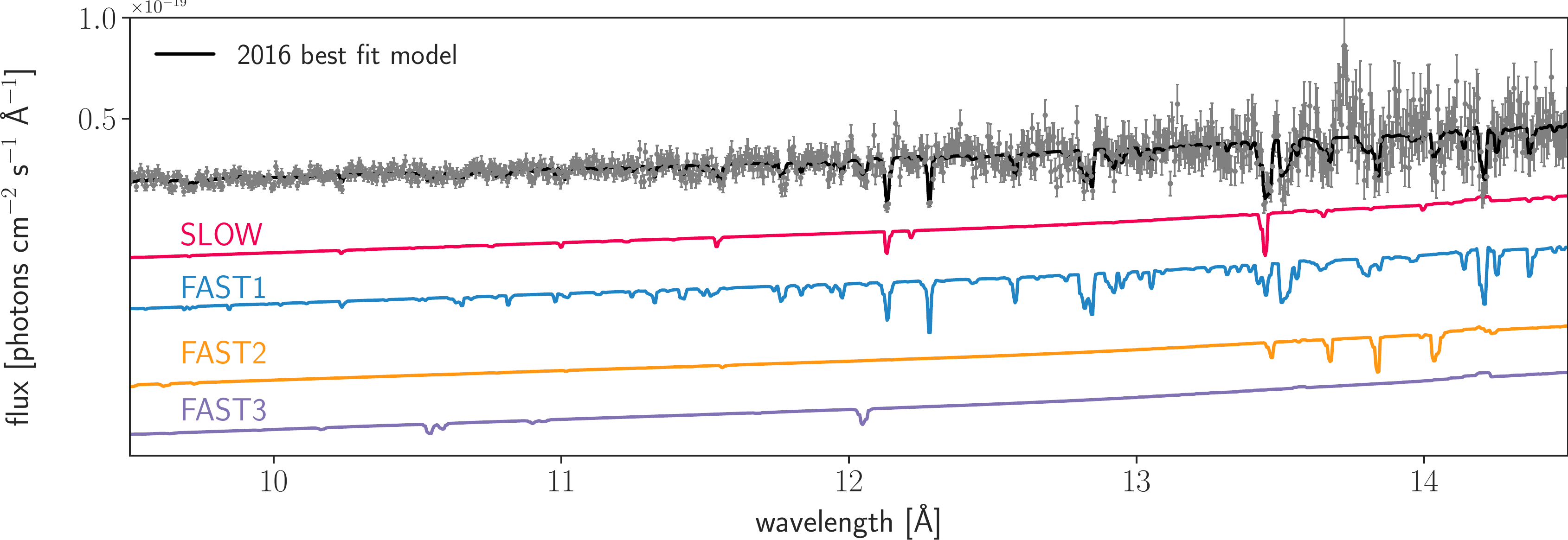}
\includegraphics[width=\columnwidth,trim={0mm 0mm 0mm 0mm},clip]{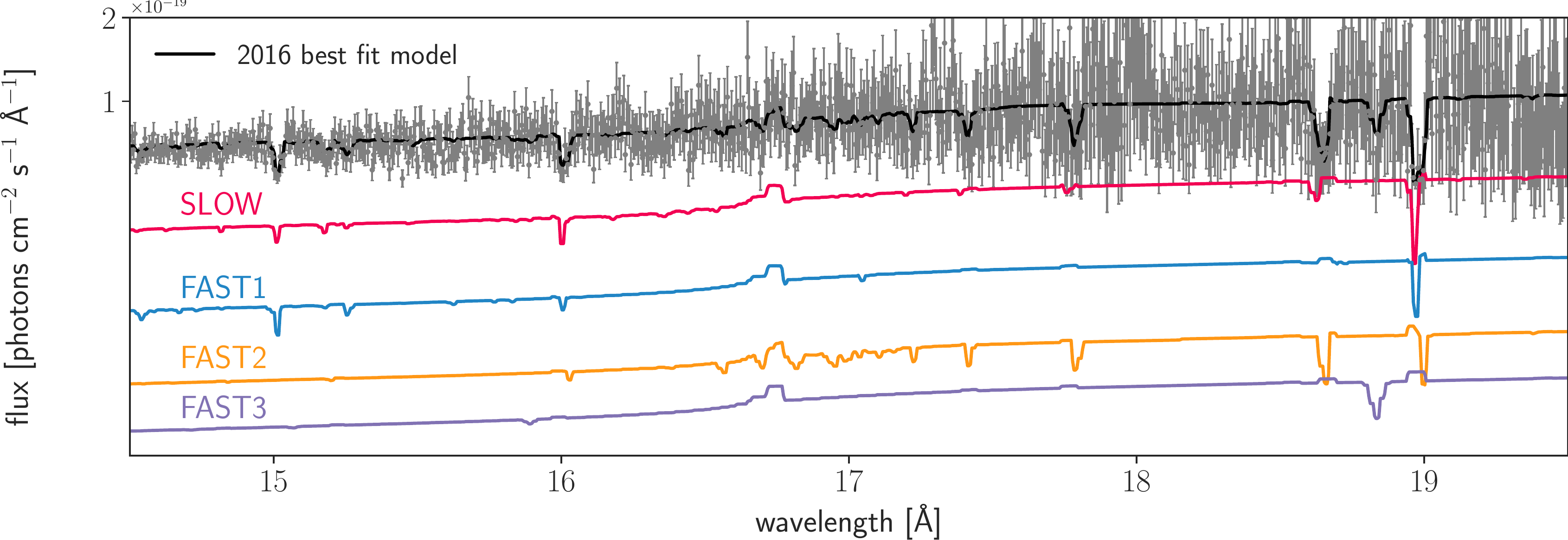}
\caption{Spectral features imprinted by narrow line width absorbers found in the 2016 data. The black line is the overall best fit model (\textit{neutral MW * hot MW * all AGN absorbers * AGN continuum}), plotted on top of the 2016 HETG/MEG data (grey). Below, narrow line width absorbers are shown individually, i.e.  \textit{neutral MW * individual absorber * AGN continuum}. For identification of the most significant individual features, please refer to Table~\ref{tab:lines16}. Several unlabelled features are from the hot halo of the Milky Way \citep[see][]{futureMW}. }
\label{fig:narrowcomps16}
\end{minipage}
\end{figure*}

\begin{table*} 
\begin{minipage}{160mm}
\centering
\caption{Absorption features between 1.5 and 20\AA\ for each wind component in the 2008 epoch. The most significant features are listed, which have the optical depths over unity across the whole line. We list the ionic specie, rest-frame energy and wavelength, and red/blueshifted wavelength for direct comparison with Figures~\ref{fig:spec08} and ~\ref{fig:narrowcomps08}-~\ref{fig:broadcomps08}.   \label{tab:lines08}}
\begin{tabular}{@{ }c@{  \: }c@{ \: }c@{ \: }c@{  \: }c@{ \: }c@{ \: }c@{  \: }c@{ \: }c@{ \: }c@{ \: }c@{ \: }c@{ \: }}
\hline
\hline
Ion & $E_{rf}$ & $\lambda_{rf}$ & $\lambda_{shifted}$  &Ion & $E_{rf}$ & $\lambda_{rf}$ & $\lambda_{shifted}$ & Ion & $E_{rf}$ & $\lambda_{rf}$ & $\lambda_{shifted}$  \\

 & [keV] & [\AA] & [\AA]  & & [keV] & [\AA] & [\AA] &  & [keV] & [\AA] & [\AA] \\

\hline
\hline
\multicolumn{12}{c}{\textit{2008/SLOW}} \\
\hline

NeVII&0.89665&13.827&13.816 & NeVI&0.88382&14.028&14.016 & NeV&0.87253&14.21&14.198 \\
NeVI&0.884&14.025&14.013 & NeVI&0.88288&14.043&14.031 & NeV&0.87125&14.231&14.219 \\

\hline
\multicolumn{12}{c}{\textit{2008/FAST1}} \\
\hline
S XV&2.462&5.0358&5.0387 & FeXX&0.96827&12.805&12.812 & FeXVIII&0.87338&14.196&14.204 \\
SiXIV&2.0066&6.1787&6.1822 & FeXX&0.96714&12.82&12.827 & FeXVIII&0.87309&14.201&14.209 \\
SiXIII&1.866&6.6442&6.648 & FeXX&0.96575&12.838&12.845 & FeXVIII&0.86311&14.365&14.373 \\
MgXII&1.4732&8.4162&8.421 & FeXIX&0.92374&13.422&13.43 & FeXVII&0.82632&15.004&15.013 \\
MgXI&1.353&9.1635&9.1687 & FeXIX&0.92194&13.448&13.456 & FeXVII&0.81284&15.253&15.262 \\
FeXXII&1.054&11.763&11.77 & FeXIX&0.91849&13.499&13.506 & O VIII&0.65399&18.958&18.969 \\
NeX&1.0224&12.127&12.134 & FeXIX&0.91723&13.517&13.525 &  &  &  \\
FeXXI&1.01&12.275&12.282 & FeXIX&0.89901&13.791&13.799 &  &  &  \\

\hline
\multicolumn{12}{c}{\textit{2008/FAST2}} \\
\hline

NeX&1.0228&12.122&12.134 & FeXVII&0.82668&14.998&15.013 & O VII&0.66629&18.608&18.627 \\
NeIX&0.92295&13.434&13.447 & O VIII&0.7754&15.99&16.006 & O VIII&0.65428&18.95&18.969 \\

\hline
\multicolumn{12}{c}{\textit{2008/FAST3}} \\
\hline

FeXXVI&8.3662&1.482&1.5027 & FeXXV&6.7943&1.8248&1.8504 & MgXII&1.493&8.3046&8.421 \\
NiXXVIII&8.2056&1.511&1.5321 & S XVI&2.6584&4.6638&4.7291 & NeX&1.0361&11.966&12.134 \\
FeXXVI&7.0637&1.7552&1.7798 & SiXIV&2.0336&6.0968&6.1822 & O VIII&0.66278&18.707&18.969 \\

\hline
\multicolumn{12}{c}{\textit{2008/VFAST1}} \\
\hline

FeXXVI&8.5568&1.4489&1.5027 & FeXXV&6.9491&1.7842&1.8504 & MgXII&1.527&8.1196&8.421 \\
NiXXVIII&8.3925&1.4773&1.5321 & S XVI&2.719&4.5599&4.7291 & NeX&1.0597&11.7&12.134 \\
FeXXVI&7.2247&1.7161&1.7798 & SiXIV&2.0799&5.961&6.1822 & O VIII&0.67788&18.29&18.969 \\

\hline
\hline
\end{tabular}
\end{minipage}
\end{table*}

\begin{table*} 
\begin{minipage}{160mm}
\centering
\caption{Absorption features between 1.5 and 20\AA\ for each wind component in the 2016 epoch. The most significant features are listed, which have the optical depths over unity across the whole line. We list the ionic specie, rest-frame energy and wavelength, and red/blueshifted wavelength for direct comparison with Figures~\ref{fig:spec16},  ~\ref{fig:narrowcomps16} and~\ref{fig:broadcomps16}. \label{tab:lines16} }
\begin{tabular}{@{ }c@{  \: }c@{ \: }c@{ \: }c@{  \: }c@{ \: }c@{ \: }c@{  \: }c@{ \: }c@{ \: }c@{ \: }c@{ \: }c@{ \: }}
\hline
\hline
Ion & $E_{rf}$ & $\lambda_{rf}$ & $\lambda_{shifted}$  &Ion & $E_{rf}$ & $\lambda_{rf}$ & $\lambda_{shifted}$ & Ion & $E_{rf}$ & $\lambda_{rf}$ & $\lambda_{shifted}$  \\

 & [keV] & [\AA] & [\AA]  & & [keV] & [\AA] & [\AA] &  & [keV] & [\AA] & [\AA] \\

\hline
\hline
\multicolumn{12}{c}{\textit{2016/SLOW}} \\
\hline
NeX&1.0226&12.125&12.134 & FeXVII&0.82648&15.001&15.013 & O VII&0.66613&18.613&18.627 \\
NeIX&0.92272&13.437&13.447 & O VIII&0.77521&15.994&16.006 & O VIII&0.65412&18.954&18.969 \\
\hline
\multicolumn{12}{c}{\textit{2016/FAST1}} \\
\hline
S XV&2.4621&5.0357&5.0387 & FeXXI&1.0101&12.275&12.282 & FeXIX&0.91726&13.517&13.525 \\
SiXIII&2.1839&5.6773&5.6807 & FeXX&0.9683&12.804&12.812 & FeXIX&0.89904&13.791&13.799 \\
SiXIV&2.0067&6.1785&6.1822 & FeXX&0.96717&12.819&12.827 & FeXVIII&0.87341&14.195&14.204 \\
SiXIII&1.8661&6.644&6.648 & FeXX&0.96579&12.838&12.845 & FeXVIII&0.87311&14.2&14.209 \\
MgXII&1.4732&8.4159&8.421 & FeXX&0.96129&12.898&12.905 & FeXVIII&0.86314&14.364&14.373 \\
MgXI&1.3531&9.1632&9.1687 & FeXIX&0.92377&13.422&13.43 & FeXVII&0.82634&15.004&15.013 \\
FeXXII&1.0541&11.762&11.77 & FeXIX&0.92198&13.448&13.456 & O VIII&0.65401&18.957&18.969 \\
NeX&1.0224&12.127&12.134 & FeXIX&0.91852&13.498&13.506 &  &  &  \\

\hline
\multicolumn{12}{c}{\textit{2016/FAST2}} \\
\hline

NeVII&0.89664&13.828&13.816 & O VII&0.69719&17.783&17.768 & O VIII&0.65305&18.985&18.969 \\
O VII&0.7121&17.411&17.396 & O VII&0.66504&18.643&18.627 &  &  &  \\

\hline
\multicolumn{12}{c}{\textit{2016/FAST3}} \\
\hline
FeXXVI&7.0205&1.766&1.7798 & SiXIV&2.0211&6.1344&6.1822 & NeX&1.0298&12.04&12.134 \\
FeXXV&6.7527&1.8361&1.8504 & MgXII&1.4838&8.3558&8.421 & O VIII&0.65872&18.822&18.969 \\
S XVI&2.6422&4.6925&4.7291 & FeXXIV&1.1767&10.537&10.619 &  &  &  \\

\hline
\multicolumn{12}{c}{\textit{2016/VFAST1}} \\
\hline

S XV&2.5345&4.8919&5.0387 & FeXXIV&1.2026&10.31&10.619 & NeX&1.0525&11.78&12.134 \\
SiXIV&2.0657&6.0021&6.1822 & FeXXIII&1.1631&10.66&10.98 & FeXXI&1.0397&11.924&12.282 \\
SiXIII&1.9209&6.4543&6.648 & FeXXIII&1.1591&10.697&11.018 & O VIII&0.67323&18.416&18.969 \\
MgXII&1.5165&8.1756&8.421 & FeXXII&1.085&11.427&11.77 &  &  &  \\

\hline
\multicolumn{12}{c}{\textit{2016/VFAST2}} \\
\hline

FeXXV&7.3297&1.6915&1.8504 & SiXIV&2.1939&5.6514&6.1822 & O VIII&0.71501&17.34&18.969 \\
S XVI&2.8679&4.3231&4.7291 & FeXXIV&1.2772&9.7072&10.619 &  &  &  \\
S XV&2.6917&4.6061&5.0387 & FeXXIII&1.2352&10.037&10.98 &  &  &  \\

\hline
\hline
\end{tabular}
\end{minipage}
\end{table*}

\bsp    
\label{lastpage}
\end{document}